\documentclass{JHEP3}
\usepackage{amsmath}
\input epsf
\usepackage{epsfig}
\usepackage{amssymb}
\usepackage{graphics}
\usepackage[active]{srcltx}
\usepackage{amsthm}
\usepackage{shuffle}

\newcommand{\bea}{\begin{eqnarray}}
\newcommand{\eea}{\end{eqnarray}}
\newcommand{\bean}{\begin{eqnarray*}}
\newcommand{\eean}{\end{eqnarray*}}
\newcommand{\nn}{\nonumber \\}

\def\O #1{\overline{#1}}

\def\W #1{\widetilde{#1}}
\def\WH #1{\widehat{#1}}

\def\eref#1{(\ref{#1})}

\def\a{{\alpha}}

\def\b{{\beta}}

\def\Label#1{\label{#1}%
  \smash{\hbox to0pt{\raise1ex\hbox{\tiny[#1]}\hss}}}

\title{One-loop CHY-Integrand of Bi-adjoint Scalar Theory}
\author{ Bo Feng$^{ab}$, Chang Hu$^{a}$
\footnote{Emails:  fengbo@zju.edu.cn, isiahalbert@126.com . The corresponding author is
Chang Hu.} \\
{$^a$\small Zhejiang Institute of Modern Physics, Zhejiang University, Zheda Road No.38, Hangzhou, 310027, P. R. China \\
$^b$ Center of Mathematical Science, Zhejiang University, Zheda Road No.38, Hangzhou, 310027, P. R. China}}

\date{\today}
\abstract{ In this paper, the one-loop CHY-integrands of bi-adjoint scalar theory has been
reinvestigated. Differing from previous constructions, we have explicitly removed contributions
from tadpole and massless bubbles when taking the forward limit of
corresponding tree-level amplitudes. The way to remove those singular contributions is to exploit the idea of "picking poles", 
which is to multiply a special  cross ratio factor with the role of  isolating terms having a particular pole structure.
}


\newpage
\begin{document}

\section{Introduction}

In recent years, a novel formulism for tree-level amplitudes of various theories has been proposed by Cachazo, He and Yuan [CHY] in a series of papers \cite{Cachazo:2013gna,Cachazo:2013hca,Cachazo:2013iea,Cachazo:2014nsa,Cachazo:2014xea}.
The formula is given  as an integral over the moduli space of Riemann spheres
\begin{equation}
    {\cal A}_n=\int \frac{(\prod_{i=1}^n dz_i)}{\text{vol}(SL(2,\mathbb{C}))}z_{ij}z_{jk}z_{ki}\prod_{a\neq i,j,k}\delta(\mathcal{E}_a)\cdot I(z_1,\cdots,z_n)
    \label{eq:CHYconstruction}
\end{equation}
where $z_i$ are puncture locations of (the) $i$-th external particles, and the denominator $\text{vol}(SL(2,\mathbb{C}))$ comes from the M\"obius invariance. i.e., the transformation $z_i\rightarrow\frac{az_i+b}{cz_i+d}$ with $ad-bc\neq 0$. The $\mathcal{E}$'s are the $scattering$ $equation$ defined as
\begin{equation}
    \mathcal{E}_a\equiv \sum_{b\neq a}\frac{s_{ab}}{z_a-z_b}=0, a=1,2,...,n
\end{equation}
with $s_{ab}\equiv(k_a+k_b)^2$ being the Mandelstam invariants. As shown in \eqref{eq:CHYconstruction}, the CHY formalism includes two parts: the integration measure with $\delta$ functions of scattering equations, which is universal for all theories, and formulating different CHY-integrands $I$ for different theories. For a theory with $n$ particles, there are $(n-3)!$ solutions to these scattering equations. A proof of this construction for bi-adjoint scalar theory and Yang-Mills theory has been provided  by Dolan and Goddard in \cite{Dolan:2013isa}. 

Working out all $(n-3)!$ solutions of scattering equation is very
burdensome even at small value of $n$. In general there is no
effective method to solve these equations analytically if $n\geq 6$.
By exploiting computational algebraic geometry method, several works
have appeared
\cite{Dolan:2014ega,Kalousios:2015fya,Huang:2015yka,Sogaard:2015dba,Dolan:2015iln,Cardona:2015eba,Cardona:2015ouc},
such as the companion matrix, the Bezoutian matrix,   the
elimination theorem, etc. An alternative way is given in
\cite{Cachazo:2015nwa,Cachazo:2019aby} by mapping the problem to the
result of bi-adjoint $\phi^3$ theory which is already known. By
applying the generalized KLT relation and Hamiltonian decomposition
of a certain $4$-regular graph, one can read out the result directly
without solving the scattering equations. In
\cite{Baadsgaard:2015voa,Baadsgaard:2015ifa,Baadsgaard:2015hia}, an
integration rule has been presented to read out Feynman diagrams for
CHY-integrands containing only single poles. For more general
CHY-integrand\footnote{A special case, i.e., Yang-Mills amplitudes,
has been discussed in \cite{Lam:2016tlk}.} with higher poles, we can
use the cross ratio identities \cite{Cardona:2016gon} to reduce the
degree of poles one by one until all poles are single poles, then
the integration rule can be applied.

Having established the  CHY formalism for tree-level amplitudes, it
is natural to ask the generalization to loop levels.  Many works
have been done in this direction
\cite{Adamo:2013tsa,Casali:2014hfa,Adamo:2015hoa,Geyer:2015bja,
Geyer:2015jch,Baadsgaard:2015hia,He:2015yua,Cachazo:2015aol,
Zlotnikov:2016wtk,Cardona:2016wcr,He:2016mzd,Gomez:2016cqb,
Gomez:2017lhy,Gomez:2017cpe,Geyer:2017ela,Ahmadiniaz:2018nvr,Agerskov:2019ryp}.
In \cite{Geyer:2015bja, Geyer:2015jch}, using the ambitwistor string
theory, through the global residue theorem, Geyer, Mason, Monteiro
and Tourkine have written down a beautiful one-loop CHY formulism
(i.e., the loop scattering equations and loop CHY-integrands)  for
various theories like super-Yang-Mills theory, n-gon, supergravity,
pure Yang-Mills and gravity theory\footnote{ Although we will focus
on one-loop case only in this paper, it is worth to mention that
using the same idea, some two loop constructions are shown in
\cite{Feng:2016nrf,Geyer:2016wjx,Geyer:2018xwu,Geyer:2019hnn}.}.
Integration rule has been generalized to loop level in
\cite{Baadsgaard:2015hia} and the CHY-integrand of bi-adjoint
scalars has been discussed. In \cite{He:2015yua},  the idea that one
loop integrands coming from the forward limit of tree-level
amplitudes has been presented,  where the bi-adjoint scalar theory
has been investigated as an example. Singular solutions related to
the forward limits have also been clarified. The role of regular and
singular solutions of scattering equations has been further
illustrated in \cite{Cachazo:2015aol}.

Above works pointed out two important issues for the one-loop
construction in the CHY frame. Firstly, when taking the forward limit,
there are singularities we should deal with carefully. Secondly, as
emphasized in \cite{Cachazo:2015aol}, when we talk about the one-loop
integrands of a given theory, we should think it in the sense of
equivalent class, i.e., expressions in the same class  can be
different by scaleless parts. In this paper, we will suggest a new
approach to treat the singularities related to the forward limit. 
The way to remove those singular contributions is to exploit the idea of "picking poles",
which is to multiply a special  cross ratio factor with the role of  isolating terms having a particular pole structure.
As
an application of our new strategy, one-loop CHY-integrands of
bi-adjoint scalar theory will be re-investigated.

The plan of the paper is follows. In the next section we provide a
lightning review of some necessary backgrounds for this paper,
 including the integration
rules that enable us to evaluate amplitude from a given
CHY-integrand, the conception and some properties of effective
Feynman diagrams, and the cross ratio factor that allow us to pick
up terms containing a given pole structure. In section three we
reinvestigate the bi-adjoint scalar theory. Firstly we review the
construction idea given in \cite{Baadsgaard:2015hia,He:2015yua} and
 the relation between loop diagrams and tree
diagrams through the cutting and gluing process. Based on the
connection, we construct the one-loop CHY-integrand for two special
cases, i.e., two PT-factors have same orderings or opposite
orderings. Combining  results of above two special cases and the
effective Feynman diagram of corresponding tree-level amplitudes, we
give a general construction of one-loop CHY-integrand for two
arbitrary PT-factors. Finally, in the section four, we give brief
summary and discussion.

\section{Backgrounds}

In this section, we will review some known results, which are useful
for our construction of one-loop CHY-integrands of bi-adjoint scalar
theory.

\subsection{Integrate rules}

In this subsection, we will review how to write down the analytic
expression for the amplitudes  given by
\begin{equation}
    A=\int d\Omega_{CHY} I(z_1,\cdots,z_n)
    ~~~\label{eq:CHY form}
\end{equation}
in the CHY formalism.  A generic CHY-integrand can be given by the
sum of following terms
\bea \mathcal{I}=\frac{1}{\prod_{1\leq i<j\leq n
}z_{ij}^{\beta_{ij}}},~~~\label{gen-term}\eea
where $z_{ij}\equiv z_i-z_j$ and $\b_{ij}$'s can be any integer
numbers\footnote{When $\b<0$, it is numerator.}. The M\"obius
invariance requires that $\sum_{j< i} \b_{ji}+\sum_{j>i}\b_{ij}=4$.
For each term like \eref{gen-term}, we can define the {\bf  pole
index} of a subset $A_i\subset \{1,2,...,n\}$ as
\bea \chi (A_i)\equiv L[A_i]-2(|A_i|-1)~,~~\label{index-1}\eea
where $|A_i|$ is the number of external particles inside the
subset $A_i$ and $L[A_i]$ is the linking number which is given by
\bea L[A_i]=\sum_{a<b;a,b\in A_i} \b_{ab}~.~~\label{index-2} \eea
It is worth to notice that the subset $A_i$ and its complement
subset $\O A_i$ have the same pole index by the M\"obius invariance
condition. When the pole index $\chi\geq 0$, the amplitude can have
terms with poles like $\frac{1}{s_{A_i}^{\chi[A_i]+1}}$, where
\bea s_{A_i}=(\sum_{a\in A_i} k_a)^2=(\sum_{b\in \O A_i}
k_b)^2~.\eea
Now we show the algorithm\footnote{There are several algorithms in
the literatures, see related references
\cite{Cachazo:2013iea,Cachazo:2015nwa,Baadsgaard:2015voa,
Baadsgaard:2015ifa,Cardona:2016gon,Baadsgaard:2015hia,Lam:2016tlk,Huang:2016zzb,
Huang:2017ydz,Huang:2018zfi,Baadsgaard:2015abc,
Bjerrum-Bohr:2016juj, Bjerrum-Bohr:2016axv}. In this paper, we
review these methods which are suitable for our purpose.
  } to read out the corresponding Feynman
diagrams from the given CHY-integrands \eref{gen-term}.

Let us start from the simplest case, i.e., the CHY-integrand $I$ is
given by the product of two Park-Taylor factors
\bea I(z_1,\cdots,z_n)=PT(\pi)\times PT(\rho)~,~~\label{two-PT} \eea
where $\pi,\rho$ are two orderings of external particles and for the
given ordering $\a$, the Park-Taylor factor is defined by
\bea PT(\pi) \equiv \frac{1}{(z_{\pi(1)}-z_{\pi(2)})(z_{\pi(2)}-z_{\pi(3)})\cdots(z_{\pi(n-1)}-z_{\pi(n)})
        (z_{\pi(n)}-z_{\pi(1)})}~~,~~~\label{PT}\eea
which is obviously cyclic invariance. For this particular case
\eref{two-PT}, \cite{Cachazo:2013iea} shows that the result is given
by the sum of the collection of all trivalent scalar diagrams that
can be regarded both as $\pi$-color-ordered and
$\rho$-color-ordered, where each diagram's contribution is given by
the product of its propagators up to a sign. To make thing clear,
let us present one example with $\pi=(12345678)$ and
$\rho=(12673458)$ as following:
\begin{figure}[ht]
    \centering
    \includegraphics[scale                                                               =0.3]{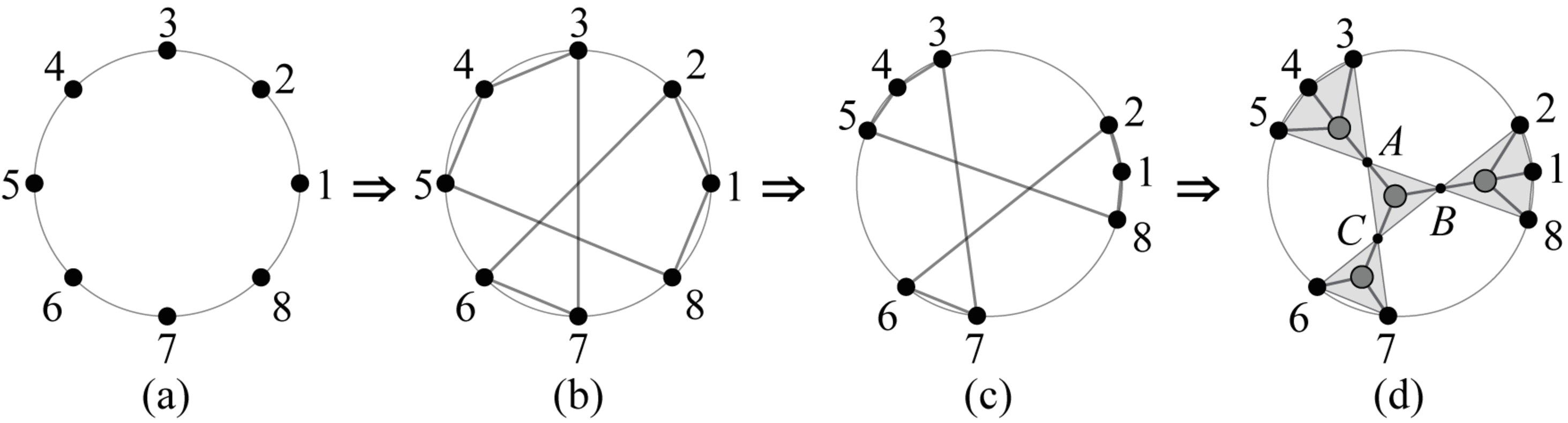}
    \caption{ The procedure to obtain the effective Feynman diagram for the ordering $\pi=(12345678)$ and $\rho=(12673458)$.}
    \label{fig:Effect12673458}
\end{figure}
\begin{itemize}
\item (a) First, we draw a  disk with $n$ nodes sitting on the boundary
in the ordering $\pi$ (see graph (a) of Figure
\underline{\ref{fig:Effect12673458}}). Then we link $n$ nodes
together with a loop of line segments according to the ordering
$\rho$ (see graph (b) of Figure
\underline{\ref{fig:Effect12673458}}). It is worth to notice
that the closed $\rho$ loop will form many small regions by its
intersections as shown in the graph (b), where there are two
pentagons, one quadrilateral and one triangle. If there are
small regions sharing  same edge, we need to pull nodes
belonging to the same region together until any two adjacent
polygons share at most one common vertex as shown in the graph
(c) of Figure \underline{\ref{fig:Effect12673458}}.
\item (b) Having obtained the valid intersection pattern as shown in graph (c),
we construct the effective Feynman diagram as following. First
we put a vertex at the middle of each small polygon (such a
vertex is called the "effective vertex"). Then we connect each
vertex with external nodes of the same polygon. Finally we
connect two effective vertexes together if their polygons share a
common point (such a line is called "effective propagator"). The
achieved figure is the effective Feynman diagram as shown in
graph (d) of Figure \underline{\ref{fig:Effect12673458}}.
\item (c) Having the effective Feynman diagram, we can easily read out
the analytic expression by noticing that each vertex with $m$
legs represents all possible Feynman diagrams with only cubic
vertexes and $m$ external legs having this ordering. For this
example, there are four vertexes. For the vertex connecting to
nodes $1,2,8$, it gives ${1\over s_{12}} +{1\over s_{18}}$.
Similar understanding for other three effective vertexes.
Multiplying contributions from these four vertexes as well as
effective propagators connecting these vertexes, we get the
final result up to a sign
\bea {1\over s_{128}} {1\over s_{345}}{1\over s_{67}}\times
 \left( {1\over s_{12}} +{1\over s_{18}}\right)\times \left(
 {1\over s_{34}}+{1\over s_{45}} \right)~.~~~\label{exa-fey}\eea
\end{itemize}
Having reviewed  the construction of effective Feynman diagrams  up
to a sign, we discuss the  overall sign given by $(-1)^{ n-3+n_{\rm
flip}(\pi|\rho)}$, where $n_{\rm flip}$ is determined as following
 \cite{Cachazo:2013iea}:
\begin{itemize}
\item (a) At the first step, we merge consecutive nodes both in the orderings
$\pi$ and $\rho$, until there is no more merging can be done.
For example, with $\pi=(123456789)$ and $\rho=(128946573)$, we
obtain $\pi[1]=(P_{12}34567P_{89})$ and
$\rho[1]=(P_{12}P_{89}46573)$.
\item (b) Next, we flip two nearby nodes in the ordering $\rho[1]$
if there two nodes are also nearby in the ordering $\pi[1]$ but
with the reversing ordering. For example, pairs $(P_{12}P_{89})$
and $(65)$ in $\rho[1]$ should be flipped. With such an
operation, we get
 contribution $(-)^{m}$ where $m$ is number of pairs doing the
 flip. With above example, we have $\pi[2]=(P_{89}
 P_{12}34567)$, $\rho[2]=(P_{89}P_{12}45673)$ and contribution
 $(-)^2$.
\item (c) Now we repeat above two steps. The merging gives $\pi[3]=(P_{8912}3P_{4567})$ and $\rho[3]=(P_{8912}P_{4567}3)$, then flipping pair $(3P_{8912} )$ gives $\pi[4]=(P_{8912}3P_{4567})$ and $\rho[4]=(P_{8912}3 P_{4567})$ with sign contribution $(-)$. Now $\pi[4]$ and $\rho[4]$ are same up to cyclic ordering and the iterating stops.
\item (d) The final sign is given by  $(-1)^{ n-3+n_{\rm flip}(\pi|\rho)}$. For our example, $n=9$ and
   $n_{\rm flip}=3$, thus the overall sign is $(-)$.
\end{itemize}
Having discussed the simplest case, i.e., the CHY-integrands given
by the product of two PT-factor, we move to the next simple case,
i.e., all poles are at most simple poles (i.e., all subsets with
index $\chi\leq 0$). For this situation, the {\bf integration rule}
has been proposed in \cite{Baadsgaard:2015voa}. To understand the
algorithm, let us use the following example to demonstrate
\begin{equation}
    I=\frac{z_{12}}{z_{13}z_{16}z_{17}z_{23}z^2_{26}z_{27}
    z_{32}z_{34}z^2_{45}z^2_{51}z_{67}z_{74}}~.~~\label{6p-exp}
\end{equation}
\begin{itemize}
\item (a) At the first step, for the given CHY-integrand \eqref{eq:CHY form},
we find all subsets $A_i$ having pole index $\chi(A_i)\geq
0$\footnote{Since from the point view of pole structures, the
subset $A_i$ is equivalent to its complement subset
$\bar{A_i}=\{1,2,\cdots,n\}-A_i$, we keep only one of them.}.
The case $\chi(A_i)=0$ corresponds the simple pole. For  our
example \eref{6p-exp}, the simple pole sets are following seven
subsets:
\bea \{1,5\},\{2,3\},\{1,4,5\},\{2,3,6\},
\{2,6\},\{4,5\},\{2,6,7\}~.~~\label{6p-exp-pole}\eea
\item (b) \textbf{Compatible Condition:} Having found all poles,
we need to find the compatible combination of subsets as large
 as possible, where each pair of subsets $\{A_i,A_j\}$ in the
 combination is either nested (i.e, $A_i\subset A_j$ or
 $A_j\subset A_i$) or complementary $A_i\bigcap A_j=\emptyset$.
 For  our example \eref{6p-exp}, by checking \eref{6p-exp-pole}
 we have following six maximum combinations:
\begin{equation}
    \begin{aligned}
    &\{\{1,5\},\{2,3\},\{1,4,5\},\{2,3,6\}\},\ \
    \{\{1,5\},\{2,6\},\{1,4,5\},\{2,3,6\}\},\\
    &\{\{2,3\},\{4,5\},\{1,4,5\},\{2,3,6\}\},\ \
    \{\{2,6\},\{4,5\},\{1,4,5\},\{2,3,6\}\},\\
    &\{\{1,5\},\{2,6\},\{1,4,5\},\{2,6,7\}\},\ \
    \{\{2,6\},\{4,5\},\{1,4,5\},\{2,6,7\}\}.~~~\label{6p-exp-pole-comb}
    \end{aligned}
\end{equation}

\item (c) \textbf{Length Condition:} For a maximum combinations to
 give a nonzero Feynman diagram,  the number of subsets in the
 maximum combination of \eref{6p-exp-pole-comb} should be
 exactly $(n-3)$.  In our example, all six maximum combinations
 satisfy the length condition.

\item (d) For each maximum compatible combination satisfying the
 length condition, we draw the corresponding Feynman diagram
 with only cubic vertices, where each  propagator corresponds to
 a pole subset. After summing up all Feynman diagrams together,
 we get the final result for the starting CHY-integrand (up to a
 sign). For our example, it is
\begin{equation}
\begin{aligned}
    &\frac{1}{s_{15}s_{23}s_{145}s_{236}}+\frac{1}{s_{15}s_{26}s_{145}s_{236}}+\frac{1}{s_{23}s_{45}s_{145}s_{236}}\\+&\frac{1}{s_{26}s_{45}s_{145}s_{236}}+\frac{1}{s_{15}s_{26}s_{145}s_{267}}+\frac{1}{s_{26}s_{45}s_{145}s_{267}}
\end{aligned}
\end{equation}

\item (e) The way to determine the overall sign for the case with simple poles is a little bit complicated
and more details can be found in
\cite{Baadsgaard:2015abc,Cardona:2016gon}.

\end{itemize}

For more general CHY-integrand with higher poles, we can use the
cross ratio identities \cite{Cardona:2016gon} to reduce the degree
of poles one by one until all poles are single poles\footnote{A
clear algorithm has been given in the Appendix of
\cite{Huang:2017ydz}. Other related understanding of higher poles
can be found in \cite{Huang:2016zzb,Zhou:2017mfj}.}, then we can use
the integration rule reviewed above to write down analytical
expressions.

\subsection{Effective Feynman diagram}

To construct one-loop CHY-integrand for a given theory, some
knowledge of the corresponding Feynman diagrams will be useful. For
the bi-adjoint scalar theory focused on in this paper, the concept
 {\bf effective Feynman diagram} \cite{Cachazo:2013iea,Huang:2018zfi}, which has already appeared
in the graph (d) of Figure \underline{\ref{fig:Effect12673458}}, becomes very
important. More explicitly, for any pair of PT factors with nonzero
contributions, all Feynman diagrams can be coded into one effective
Feynman diagram as explained in \eref{exa-fey}. Another example with
$[\pi|\rho]\equiv [12345678|12846573]$, where two orderings (up to
cyclic permutations) have been separated by the vertical line, will
give the effective Feynman diagram in \eref{eq:effect12846573}
\begin{equation}
    \includegraphics[scale=0.43]{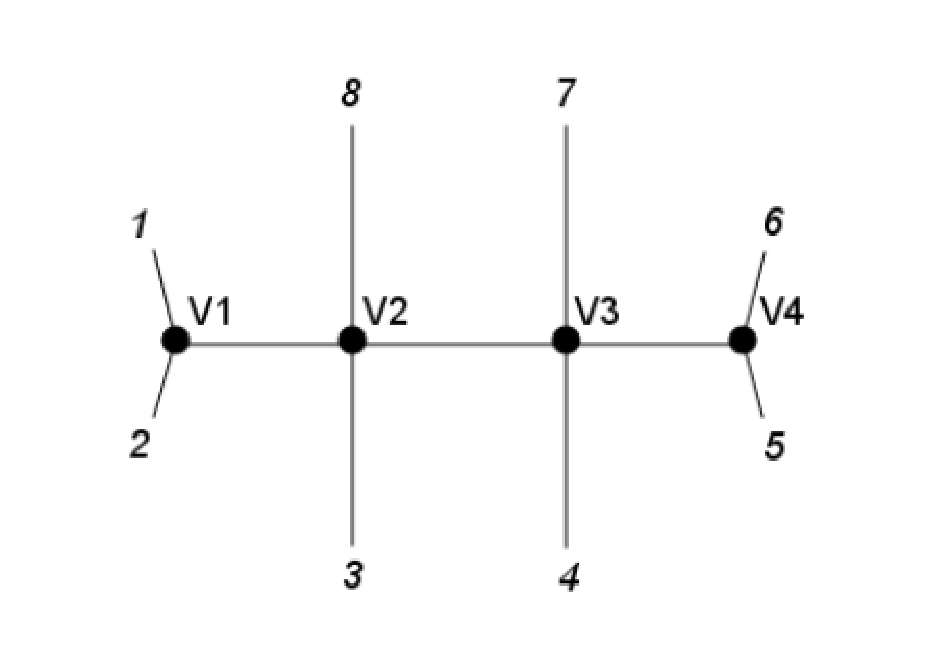}
    \label{eq:effect12846573}
\end{equation}
Given the effective Feynman diagram, we can infer following information:
\begin{itemize}

\item (1) For each effective vertex, it will split each ordering into
several groups.  For example,  for vertex $V_2$ in
\eqref{eq:effect12846573},  $\pi$-planar ordering
($\pi=[12345678]$) is split into $(12)(3)(4567)(8)$, while
$\rho$-planar ordering, to $(12)(8)(4657)(3)$.

\item (2) If we treat each group as an element for the splitting of a given effective vertex,
 the relative ordering between $\pi$- and $\rho$-planar
 orderings must be either same or complete reversed. For
 example, for $V_1$, we have  the $\pi$-planar ordering
 splitting $(1)(2)(345678)\equiv (1)(2)(P)$ and the
 $\rho$-planar ordering splitting  $(1)(2)(846573)\equiv
 (1)(2)(P)$, thus they are in same ordering.  For $V_2$, we have
 the $\pi$-planar ordering splitting  $(12)(3)(4567)(8)\equiv
 (P_1)(3)(P_2)(8)$ and $\rho$-planar ordering splitting
 $(12)(8)(4657)(3)\equiv (P_1)(8)(P_2)(3)$, thus they are in
 reversed ordering.
 For simplicity and clarity we
 call the former one $\cal{SO}$-type (abbreviation of 'same
 ordering') and the latter one $\cal{RO}$-type (abbreviation of
 'reversed ordering').

\item (3) Another fact is that the relative ordering of two adjacent
 effective vertices must be opposite. In other words, one of
 them is $\cal{SO}$-type and another is $\cal{RO}$-type.

\item (4)  These two properties can also be seen from the process
 of obtaining effective Feynman diagram (as explained in the
 Figure \underline{\ref{fig:Effect12673458}}). Once we fixed the planar
 order according to $\pi$, the order inside one polygon is
 determined by $\rho$, which will be either clockwise or
 anti-clockwise as shown in the graph A of Figure
 \ref{fig:2properties}, and two adjacent polygons are always in
 reversed orderings (clockwise versus anti-clockwise), as  shown
 in the graph B of Figure \ref{fig:2properties}.

\begin{figure}[ht]
    \centering
    \includegraphics[scale=0.4]{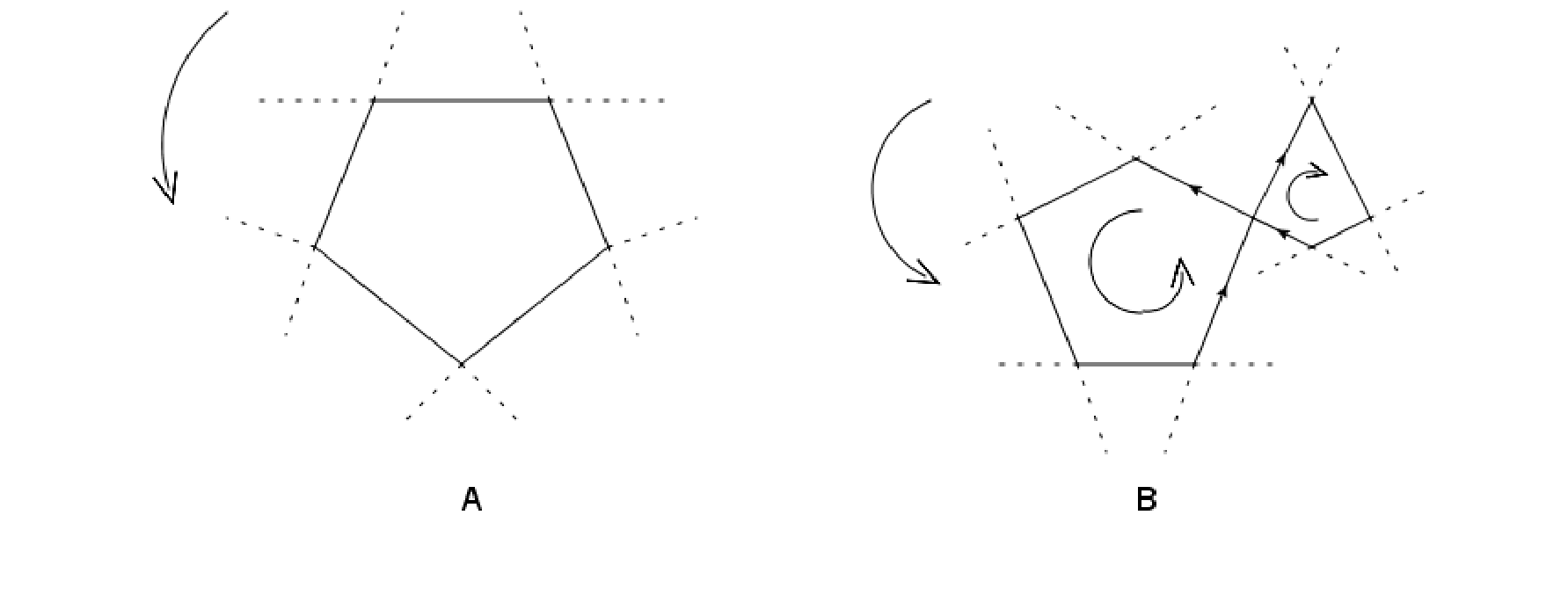}
    \caption{The pilot process shows two properties of effective Feynman diagram}
    \label{fig:2properties}
\end{figure}

\end{itemize}

\begin{figure}[ht]
    \centering
    \includegraphics[scale=0.25]{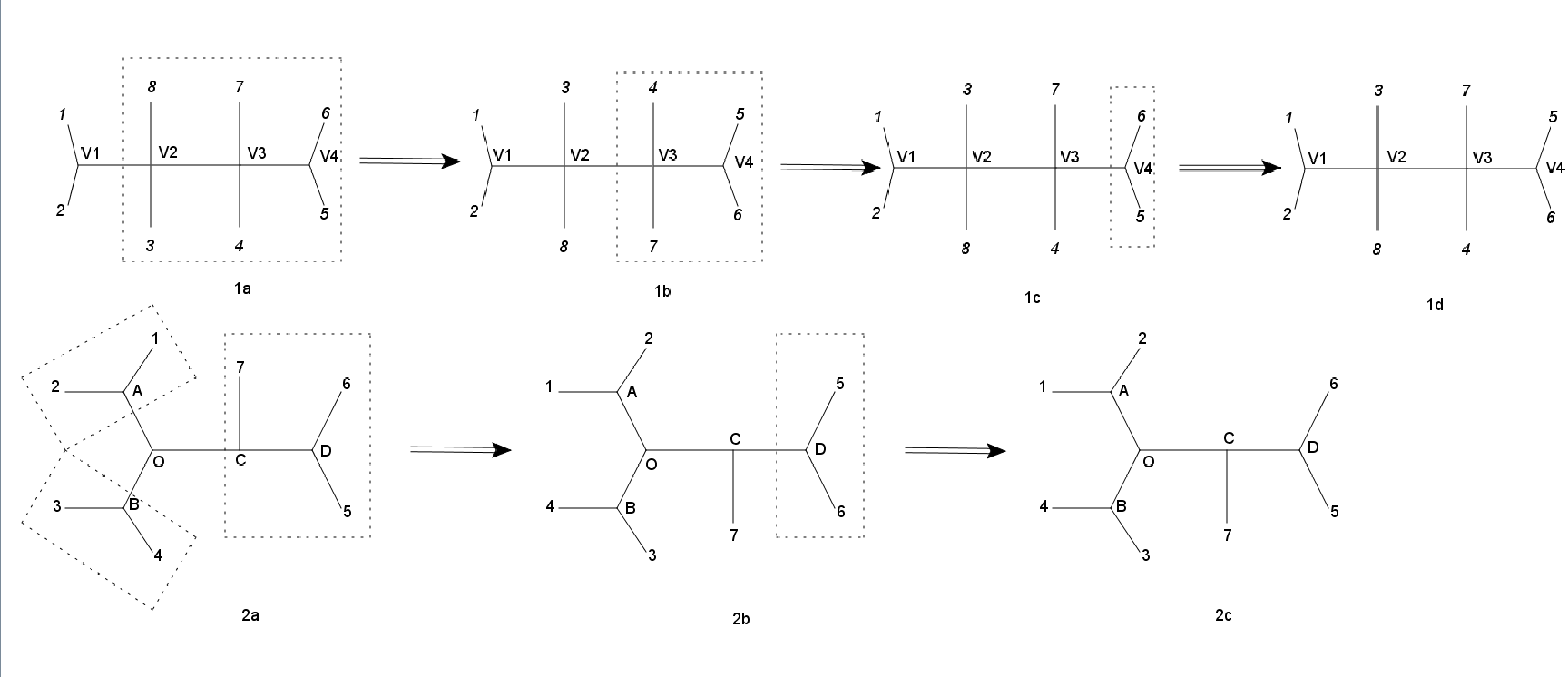}
    \caption{The process from $\pi$-ordering effective Feynman diagram to $\rho$-ordering}
    \label{fig:process-pi-to-rho}
\end{figure}

Having the effective Feynman diagram drawn in the $\pi$-planar
ordering, we can reach the $\rho$-planar ordering, by reversing
permutation mentioned in \cite{Huang:2018zfi}. Here we provide a
pictorial manipulation over the effective Feynman diagram, which
will be used for later discussion. Let us use two examples, i.e.,
$[12345678|12846573]$ and $[1234567|2143756]$, to demonstrate the
idea (see Figure \ref{fig:process-pi-to-rho}). At the first step, we
pick up an arbitrary $\cal{SO}$-type effective vertex. Then for each
sub-part connecting to this vertex through an effective propagator,
we do the axial reflection according to this propagator. After that,
we move to next effective vertex, and do the axial reflection to
remaining parts. Repeating this procedure, finally we will end up
with a $\rho$-planar ordering for the effective Feynman diagram. Let
us use the upper part of Figure \ref{fig:process-pi-to-rho} to
demonstrate the algorithm. We start with $V_1$ vertex. Among the
three parts, only the part $(345678)$ is nontrivial under the
reflection. After doing the axial reflection, we reach the graph
(1b). Next, we move to the vertex $V_2$. Among these  four parts,
the three parts $(3), (8), (4567)$ are remaining part from the point
of view of $V_1$, thus we need to do the axial reflection for them
(although for the part $(3), (8)$ it is trivial) to reach the graph
(1c). Repeating the procedure to $V_3$, finally we reach the graph
(1d), which is in the $\rho$-planar ordering.

\subsection{Picking out poles}
In \cite{Feng:2016nrf}, to pick terms containing a particular pole structure coming from CHY-integrands, such as $ (PT(1,2\cdots))^2$, following cross ratio factor
\begin{equation}
{\cal P}^{a,b}_{c,d}\equiv {z_{ab} z_{cd} \over z_{ad} z_{cb}}.~~~
  \label{eq:pinching operator}
\end{equation}
has been introduced. For example, if we want to pick terms
containing pole $s_{12}$, we do the following observation.
The pole structure has split the $n$ nodes into two subsets:
$A=\{1,2\}$ and $\O A=\{3,...,n\}$. Among these two subsets, there
are two connection lines $\{2,3\}$ and $\{1,n\}$\footnote{ If there
is a factor $z_{ab}$ (no matter which power) in the denominator of a
given CHY-integrand, where $a\in A$ and $b\in \O A$, we will call
there is a connection line between the subset $A$ and its complement
subset $\O A$. For convenience, when we write the pair of nodes in
the superscript or subscript of $\cal P$, the first is always
belonging to the subset $A$ and the second one, $\O A$.}. Using the
two pairs, we can construct the cross ratio factor ${\cal
P}^{2,3}_{1,n}$ (see  \eref{eq:pinching operator}), then one can
check that by multiplying this cross ratio factor, we get the
CHY-integrand $ (PT(1,2\cdots))^2{\cal P}^{2,3}_{1,n}$, which will
produce only these Feynman diagrams containing the pole $s_{12}$ in
the original theory. Because of this role, we will call multiplying the
factor ${\cal P}^{2,3}_{1,n}$ as "picking out the pole $s_{12}$".

We can pick up more than one pole at the same time. For example, to
pick out poles $s_{n12}$ and $s_{456}$  for $ (PT(1,2\cdots))^2$, we
use the cross ratio factor ${\cal P}^{6,7}_{4,3}$ to pick out the
pole $s_{456}$ and the cross ratio factor ${\cal P}^{n,n-1}_{2,3}$
to pick out the pole $s_{n12}$. When multiplying these two together,
the CHY-integrand $ (PT(1,2\cdots))^2{\cal P}^{6,7}_{4,3}{\cal
P}^{n,n-1}_{2,3}$ does pick out all Feynman diagrams containing the
pole structure $s_{456}s_{n12}$. Similarly, the  CHY-integrand $
(PT(1,2\cdots))^2{\cal P}^{2,3}_{1,n}{\cal P}^{1,n}_{3,4}$ does pick
out all Feynman diagrams containing the pole structure
$s_{12}s_{123}$.

The poles picking  through some cross ratio factors has been
generalized to  the CHY-integrands $PT(\a)PT(\b)$ in
\cite{Huang:2018zfi}, where there are more connection lines between
the two subsets $A$ and $\O A$ since each PT-factor can give two
lines. For this special case, as shown in \cite{Huang:2018zfi}, one
can use either PT-factor to define the corresponding cross ratio
factor, both of them give the same result.  For examples, the
CHY-integrand $PT(1,2,3,4,5,6)PT(1,3,4,6,5,2)$ gives
$-\frac{1}{s_{12}s_{34}s_{56}}-\frac{1}{s_{12}s_{56}s_{123}}$. To
pick up the pole $s_{34}$, we can use either the $PT(1,2,3,4,5,6)$
to define the cross ratio factor ${\cal P}^{4,5}_{3,2}$ or the
$PT(1,3,4,6,5,2)$ to define the cross ratio factor ${\cal
P}^{4,6}_{3,1}$. One can check that both of them give the same
results:
\bea
    && \int d\Omega_{CHY}PT(1,2,3,4,5,6)PT(1,3,4,6,5,2){\cal P}^{4,5}_{3,2}
    =-\frac{1}{s_{12}s_{34}s_{56}} \nn
    & &\int d\Omega_{CHY}PT(1,2,3,4,5,6)PT(1,3,4,6,5,2){\cal P}^{4,6}_{3,1}=-\frac{1}{s_{12}s_{34}s_{56}}
        ~.~~\label{exam-t-1}
\eea
There are two remarks about the procedure of picking out poles.
Firstly, the procedure will not change the overall sign in final
result. For example, the result \eref{exam-t-1} has the minus sign
which is the same as  the one in the original theory. Secondly, as
pointed out in \cite{Huang:2018zfi,Feng:2016nrf},  there are two
special cases where picking out procedure will fall: one is that
there is no such a pole and another one, all terms have such a pole.
For example, the CHY-integrand  $PT(1,2,3,4,5)PT(1,3,2,4,5)$ gives
$\frac{1}{s_{23}s_{15}}+\frac{1}{s_{23}s_{45}}$. If we naively pick
out the pole $s_{34}$ from the factor $PT(1,2,3,4,5)$, we will find
\bea  \int d\Omega_{CHY}PT(1,2,3,4,5)PT(1,3,2,4,5){\cal
P}^{4,5}_{3,2}=\frac{1}{s_{15}s_{24}} ~~~~\label{exception1}\eea
which is nonsense result. For the same integrand, if we try to pick
out the common pole $s_{23}$ naively, we will find
\bea \int d\Omega_{CHY}PT(1,2,3,4,5)PT(1,3,2,4,5) {\cal
P}^{2,1}_{3,4}=\frac{1}{s_{15}s_{24}}+\frac{1}{s_{15}s_{24}}
+\frac{1}{s_{13}s_{24}}+\frac{1}{s_{13}s_{45}}+\frac{1}{s_{23}s_{45}}
~,~~  \label{exception2}\eea
which has additional terms comparing to the original theory. In the
later part of the paper, when we apply the picking pole procedure,
we need to be careful about the two special cases.


\section{One loop}

In this section, we will construct the CHY-integrands for one loop amplitudes of bi-adjoint scalar theory. The basic strategy is based on following observation: by cutting a propagator containing the loop momentum, one-loop
Feynman diagram becomes the tree-level Feynman diagram. Reversing the process, i.e., by gluing two external legs
of a tree-level Feynman diagram, we get an one-loop Feynman diagram, if the momenta of these two legs are
massive and opposite, i.e., $k_{\pm}=\pm L$, $L^2\neq 0$. Such a strategy has been named the "forward limit",
which has been used in the literatures to construct CHY-integrands of various theories.


To see more clearly the connection of forward limit of $(n+2)$-point tree level amplitudes with two massive particles and  one-loop integrand of $n$ massless particles, let us consider a typical term of one loop integrands with loop momentum $L$~\footnote{These factors do not contain the loop momentum $L$ have been omitted in \eref{eq:only loop}.  }
\begin{equation}
    I(L,K_1,...,K_m)=\frac{1}{L^2(L-K_1)^2(L-K_1-K_2)^2\cdots(L-K_1-\cdots-K_{m-1})^2},
    \label{eq:only loop}
\end{equation}
where the $K_i$ is the $i$-th  momentum attached to the loop. Applying  the partial fraction identity
\begin{equation}
    \frac{1}{\prod D_i}=\sum_{i}\frac{1}{D_i\prod_{j\neq i}(D_j-D_i)},
    \label{eq:partial fraction}
\end{equation}
the \eref{eq:only loop} can be split to $m$ terms. For the $i$-th term with the propagator $D_i$, we can shift the loop momentum to $\W L\equiv L-\sum_{t=1}^i K_t$, and the $i$-th term becomes ${1\over \W L^2} {1\over \prod_{j\neq i} (D_j-D_i)}$. Since the loop integration is invariant under the shifting of loop momentum, we can sum them up to get a new representation of the loop integrand
\begin{equation}
    I(L,K_1,\cdot,K_m)\sim \frac{1}{L^2}\sum^{m}_{a=1}\prod^{a+m-2}_{b=a}\frac{1}{(L-K_a-K_{a+1}-\cdots-K_b)^2-L^2}
    \label{eq:rewrite only loop}
\end{equation}
where those labels are defined by module $m$. On the right hand side of \eqref{eq:rewrite only loop},
after pulling out the universal factor ${1\over L^2}$, each remaining part can be interpreted as a ladder-like tree diagram with two external legs having on-shell massive momenta $+L$ and $-L$ momenta (i.e., the forward limit). With this understanding the \eqref{eq:rewrite only loop} can be represented pictorially as
\begin{equation}
    \centering
    \includegraphics[scale=0.3]{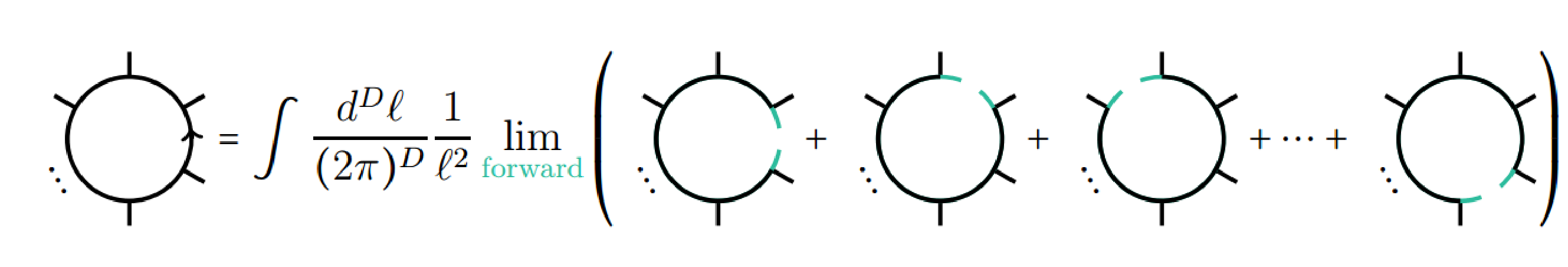}
    \label{eq:loop expand to tree}
\end{equation}
where each term inside the bracket of Right-Hand-Side of \eqref{eq:loop expand to tree} can be obtained by cutting every loop propagator. Equation \eqref{eq:loop expand to tree} gives the essential relation between tree-level and one-loop CHY-integrands, which will be the central object of  this paper.

Based on this picture, for example in \cite{Baadsgaard:2015hia,He:2015yua,Cachazo:2015aol}, a construction  of one-loop amplitudes of bi-adjoint scalar theory by taking the forward limit of tree-level amplitudes of $n$ massless and two massive particles (or two massless particle in higher dimension) has been proposed. For general
cases the expression
\begin{equation}
   m^{1-loop}_{n}[\pi|\rho]=\int\frac{d^{D}l}{(2\pi)^D}\frac{1}{l^2}\lim_{k_{\pm}\to\pm l}\sum_{\substack{\alpha\in cyc(\pi)\\\beta\in cyc(\rho)}}m^{tree}_{n+2} [-\alpha+|-\beta+]~,
    \label{eq:HY loop to tree}
\end{equation}
has been given in \cite{He:2015yua},
where
\begin{equation}
    m^{tree}_{n+2}[-\alpha+|-\beta+]=\int d\Omega_{CHY}PT(-,\alpha(1),\cdots,\alpha(n),+)PT(-,\beta(1),\cdots,\beta(n),+)~~~\label{HY-tree}
\end{equation}
is the tree-level amplitudes of $(n+2)$ particles defined by the
CHY-integrand  of two PT factors. In \eref{HY-tree}, $+,-$ denote
two extra  massive particles which can be treated as loop momentum
after taking forward limit.

Now we discuss some technical and subtle details. The pictorial
representation given in \eref{eq:loop expand to tree} has, in fact,
not one-to-one correspondence. The standard one-loop Feynman
diagrams at the LHS of \eref{eq:loop expand to tree} should be the
connected and amputated diagrams while the RHS contains, in the
principle, tadpoles and massless bubbles  after gluing\footnote{In
the formula \eref{eq:HY loop to tree}, tadpoles have been removed,
but the massless bubbles are left.}. These tadpole and massless
bubble diagrams are singular under the forward limit, thus one needs
to deal with them carefully under the limit. As shown in
\cite{He:2015yua}, we should pay attention to the relative ordering
between following two processes: solving the tree-level scattering
equations and taking the forward limit. According to the different
orderings, we can divide them to following three limits:
\begin{itemize}

\item (A) If we solve the tree-level scattering
equations first, we will have $(n-1)!$ solutions. Among these
solutions, $(n-1)!-2(n-2)!$'s are regular, while $(n-2)!$'s are
Singular I and another $(n-2)!$'s are Singular II. After summing
over all solutions for a given CHY-integrand, we take the
forward limit $k_\pm \to \pm L$.

\item (B) Without gauge fixing the scattering equations of
$k_\pm$, we take the forward limit over the tree-level
scattering equations first. Then we solve the resulted one-loop
scattering equations. We will have $(n-1)!-(n-2)!$ solutions,
which include the regular one and Singular I. Finally we sum
them over for a given CHY-integrand.

\item (C) After gauge fixing the scattering equations of
$k_\pm$, we take the forward limit over the tree-level
scattering equations first. Then we solve the resulted one-loop
scattering equations. We will have $(n-1)!-2(n-2)!$ solutions,
which include only the regular one. Finally we sum them over for
a given CHY-integrand.

\end{itemize}
It is clear that if the
CHY-integrands behave good enough, i.e., singular solutions give
zero contributions, the difference among these three categories, i.e., the contributions of Singular I and Singular II solutions are all same. But in general, singular
solutions do give nonzero contributions, thus three different
orderings of limits will lead to different expressions. However,
different expressions do not necessarily mean contradiction in the
theory. As emphasized in \cite{Cachazo:2015aol}, we should treat
loop integrands as "equivalent classes" modulo integrals that are
scaleless (thus their loop integrations are zero). In fact,
\cite{Cachazo:2015aol} pointed out that as long as CHY-integrand is
homogeneous in loop momentum $L$ under the forward limit, singular
solutions contribute only to scaleless terms, thus choosing the
Limit (C) provides a natural regularization. The theory we are
focusing in this paper, i.e., the bi-adjoint scalar theory, belongs
to the homogeneous category discussed in \cite{Cachazo:2015aol},
thus using only regular solutions should be a good practice,
especially doing the numerical calculations. However, if we want to
find the analytic expression, as carefully analyzed in
\cite{Cachazo:2015aol}, summing only over regular solutions will
produce unphysical poles. Unlike the tree-level case, we have
various ways (such as the integration rule reviewed in previous
section) to write down analytic expressions. To surround this
difficulty, in this paper we will take the different approach, i.e.,
using the Limit (A).

Our new approach, i.e., using the Limit (A), has its advantage and
disadvantage when including the singular solutions. On the one hand,
since we have summed over all solutions, methods like the
integration rule can be applied directly. On the other hand, diagrams
like tadpoles and massless bubble will cause singular behavior under
the forward limits, thus we need to treat them carefully. Our
treatment is to subtract these singular terms directly from the
CHY-integrands \eref{eq:HY loop to tree} by using the picking out
poles procedure reviewed in the previous section.


Having discussed the idea, the structure of this section is
following. In the first subsection, we discuss the case where two
orderings $\pi, \rho$ are same. In the second subsection, we discuss
the case where two orderings $\pi, \rho$ are opposite to each other.
In the third subsection, we discuss the general case. In the fourth
subsection, we present some examples.

\subsection{$\pi=\rho$}

\begin{figure}[ht]
    \centering
    \includegraphics[scale=0.27]{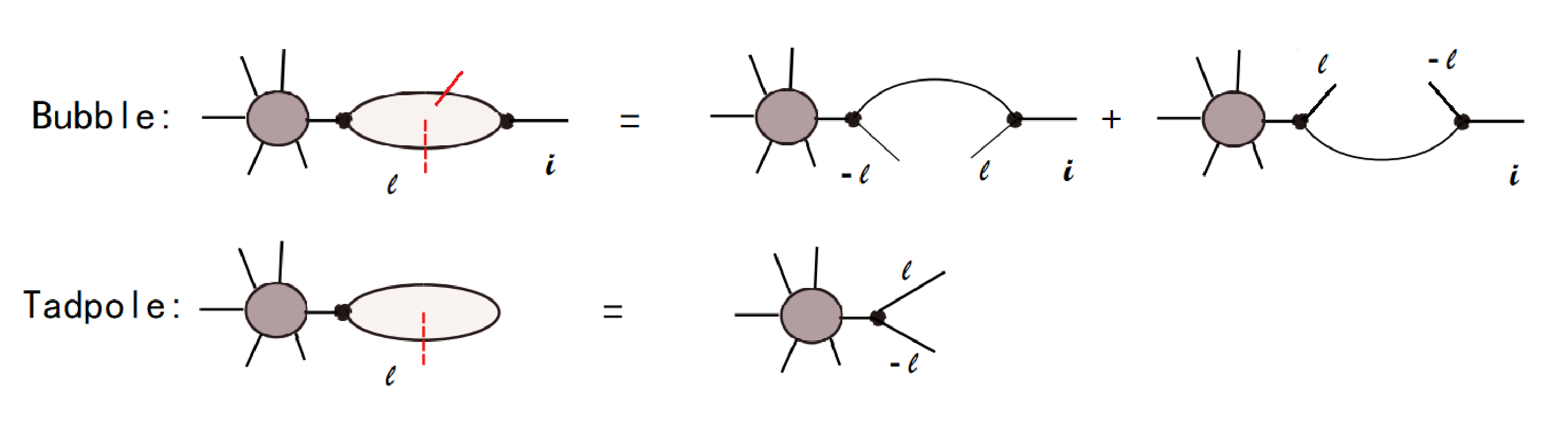}
    \caption{The should-be-excluded one-loop Feynman diagrams of $\phi^3$ theory
    and their corresponding tree diagrams after cutting}
    \label{fig:tadpole and bubble}
\end{figure}

In this subsection, we will focus on the special case $\pi=\rho$, so
$m^{1-loop}_n[\pi|\pi]$ will be the sum of all possible $n$-point
amputated one loop Feynman diagrams  with the $\pi$-planar ordering.
Without loss of generality, we can choose $\pi=(1,2,\cdots,n)$.

To get some sense for possible one-loop CHY-integrand, we need to
capture relations between one-loop diagrams and tree diagrams. These
relations can be seen by two types of processes as follows.
Recalling the expansion \eqref{eq:loop expand to tree}, we see terms
in the RHS of \eqref{eq:loop expand to tree} are the ladder-like
$(n+2)$-point tree  diagrams obtained by cutting each propagator of
loop circle in the LHS of \eqref{eq:loop expand to tree} (we call it
"cutting" process). With the ordering of external particles as
$(1,2,\cdots,n)$, we could see the orderings of particles in those
tree diagrams is $(1,2,\cdots,i,+,-,i+1,\cdots,n)$, which is
obtained by inserting the pair $(+,-)$ into two adjacent legs
$(i,i+1)$ with $i=1,2,...,n$. On the other hand, given an arbitrary
$(n+2)$-point planar tree diagram with the planar ordering
$(1,2,\cdots,i,+,-,i+1,\cdots,n)$, we can obtain an one-loop diagram
with the planar ordering $(1,2,\cdots,n)$ by linking the two
external legs $+$ and $-$ (we denote this procedure as "gluing"
process).

With above cutting and gluing processes, one natural question is
that after cutting process will two different one-loop diagrams
produce a same $(n+2)$-point planar tree diagram? Apparently the
answer is no, because by the gluing process, a $(n+2)$-point planar
tree diagram will produce one and only one one-loop diagram. This
argument shows that the cutting process and the gluing process, are
inverse operation to each other. Thus we have the  identity:
\begin{equation}
    \W M^{1-loop}_n[12\cdots n]=\int\frac{d^Dl}{(2\pi)^D}\frac{1}{l^2}\lim
    \sum_{1\leq i\leq n}m^{tree}_{n+2}[1,2,\cdots,i,+,-,i+1,\cdots,n|1,2,\cdots,i,+,-,i+1,\cdots,n]
    \label{eq:HY}
\end{equation}
where we have used the notation $\W M^{1-loop}_n[12\cdots n]$
instead of $m^{1-loop}_n[12\cdots n]$. The reason is that when
gluing tree diagrams at the RHS of \eref{eq:HY}, it will produce the
one-loop tadpole and massless bubble diagrams (see Figure
\ref{fig:tadpole and bubble}), which should not be included  for the
standard one-loop amplitude $m^{1-loop}_n[12\cdots n]$.

Now it is clear that to obtain the wanted one-loop amplitude
$m^{1-loop}_n[12\cdots n]$, we need to subtract tree diagrams from
the RHS of  \eref{eq:HY}, which will produce tadpole and massless
bubble diagrams after gluing process. These tree diagrams have
special pole structures.  From the Figure \ref{fig:tadpole and
bubble}, we see that if the $(n+2)$-point tree diagram contains the
pole $s_{+-}$, it will produce the tadpole, while if it contains the
pole $s_{i+}s_{i+-}$ or $s_{-i}s_{+-i}$, it will produce the
massless bubble of the leg $i$. Knowing the special pole structures,
the procedure of 'picking out poles' reviewed in the previous
section come to play. To isolate tree diagrams contributing to
tadpoles, we multiply the factor $ {\cal P}^{+,i}_{-,i+1}$ to the
CHY-integrand $(PT(1,2,...,i,+,-,i+1,...,n))^2$. To isolate tree
diagrams contributing to massless bubbles, we multiply the factor
${\cal P}^{i,i-1}_{+,-}{\cal P}^{i,i-1}_{-,i+1}$ to the
CHY-integrand $(PT(1,2,...,i,+,-,i+1,...,n))^2$ for massless bubble
on the leg $i$ and the factor ${\cal P}^{-,+}_{i+1,i+2}{\cal
P}^{+,i}_{i+1,i+2}$ to the CHY-integrand
$(PT(1,2,...,i,+,-,i+1,...,n))^2$ for massless bubble on the leg
$(i+1)$.  Subtracting them, we arrive the following one loop
CHY-integrand\footnote{In \cite{Baadsgaard:2015hia}, CHY-integrand
for the case $\pi=\rho$ has also been constructed, where the
tadpoles and massless bubbles have not been subtracted. However,
with the gauge fixing of $z_+,z_-$ to $0, \infty$ and the proper
regularization, these singular contributions have been controlled. }
\begin{equation}
    \begin{aligned}
       &m^{1-loop}_n[1,2,\cdots ,n|1,2,\cdots, n]\\
       =&\int\frac{d^Dl}{(2\pi)^D}\frac{1}{l^2}\lim_{k_{\pm}\mapsto\pm l}\int d\Omega_{CHY}
\sum_{i=1}^n (PT(1,2,\cdots,i,+,-,i+1,\cdots,n))^2\\
\times& \left\{1-
      {\cal P}^{+,i}_{-,i+1}-{\cal P}^{i,i-1}_{+,-}{\cal P}^{i,i-1}_{-,i+1}-{\cal P}^{-,+}_{i+1,i+2}{\cal P}^{+,i}_{i+1,i+2}\right\}~~~
      \label{eq:formula}
    \end{aligned}
\end{equation}
with all labels in ${\cal P}$ are understood as mod $n$.

Now we present an example to demonstrate our result \eqref{eq:formula} and compare it with the one given by \eref{eq:HY loop to tree}. For $n=4$, a typical term after cutting is the tree-level Feynman amplitude  $m^{tree}_6[1,2,3,4,+,-|1,2,3,4,+,-]$. It contains following $14$ diagrams
\begin{equation}
   \centering
   \includegraphics[scale=0.2]{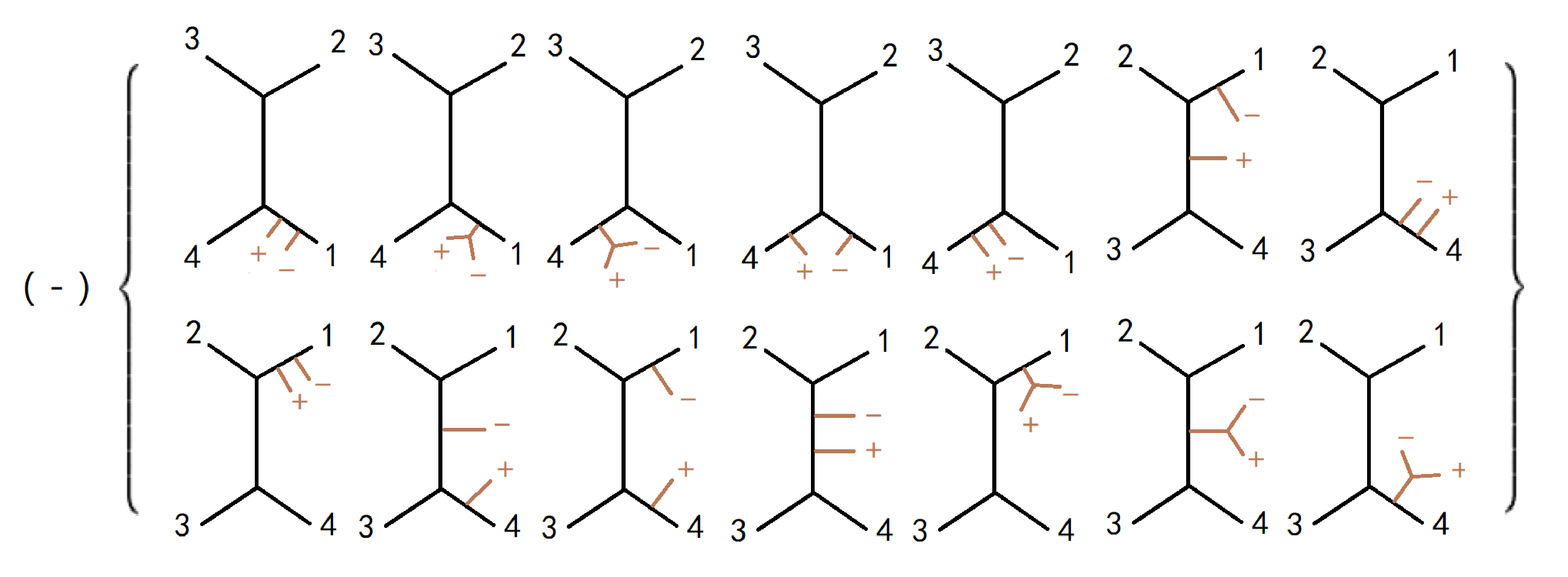}.
   \label{eq:1234-1234}
\end{equation}
Among these $14$ diagrams, five contain the pole  $s_{+-}$ and it is
easy to  check by explicit calculation that
\begin{equation}
\begin{aligned}
    \centering
    &\int d\Omega_{CHY}{\cal P}^{+,4}_{-,1}(PT(1,2,3,4,+,-))^2\\
    &\includegraphics[scale=0.2]{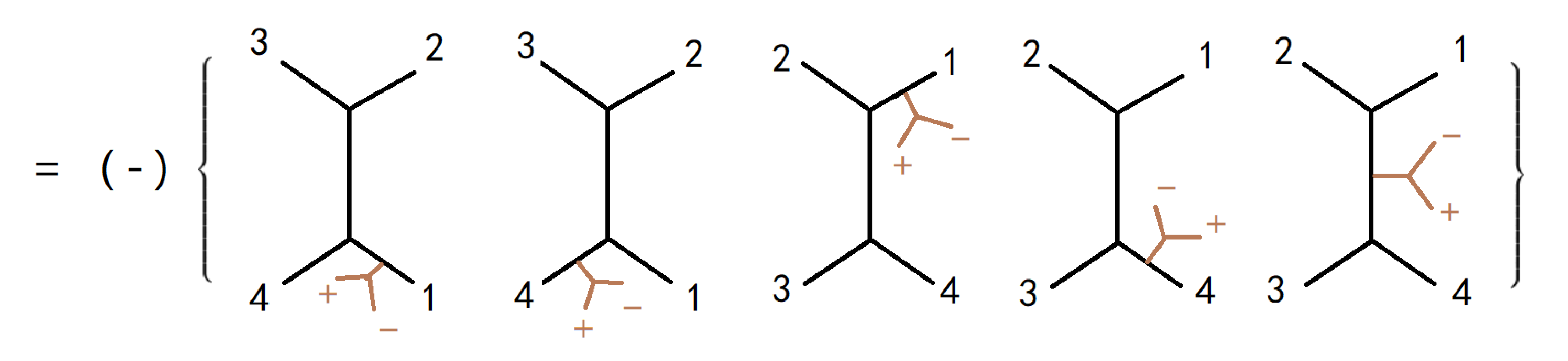}.
    \label{eq:1234tadpole}
\end{aligned}
\end{equation}
Similarly, one can check
\begin{equation}
\begin{aligned}
    \centering
    &\int d\Omega_{CHY}{\cal P}^{4,3}_{+,-}{\cal P}^{4,3}_{-,1}(PT(1,2,3,4,+,-))^2\\
    &\includegraphics[scale=0.2]{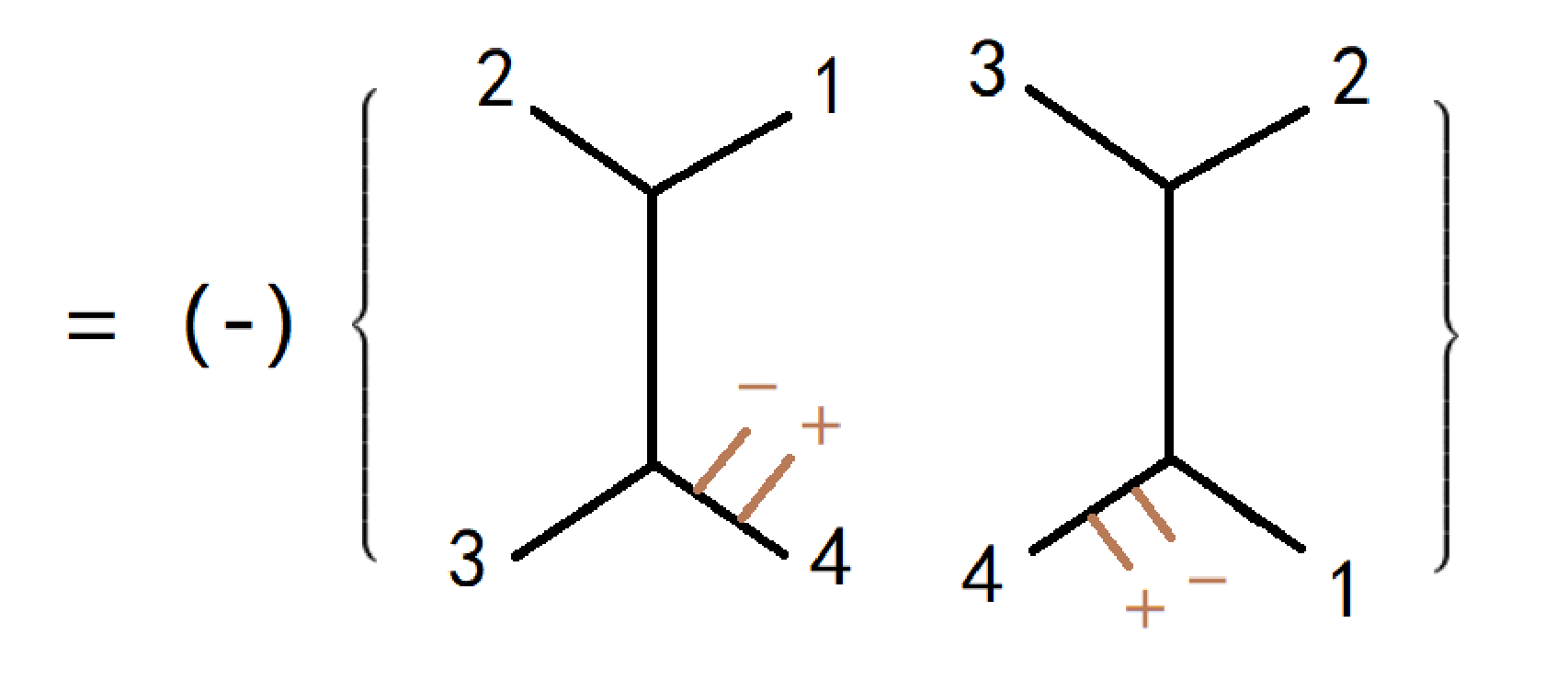}.
    \label{eq:1234bubble1}
\end{aligned}
\end{equation}
picking out all diagrams containing poles $s_{4+-}s_{4+}$, and
\begin{equation}
\begin{aligned}
    \centering
     &\int d\Omega_{CHY}{\cal P}^{-,+}_{1,2}{\cal P}^{+,4}_{1,2}(PT(1,2,3,4,+,-))^2\\
    &\includegraphics[scale=0.2]{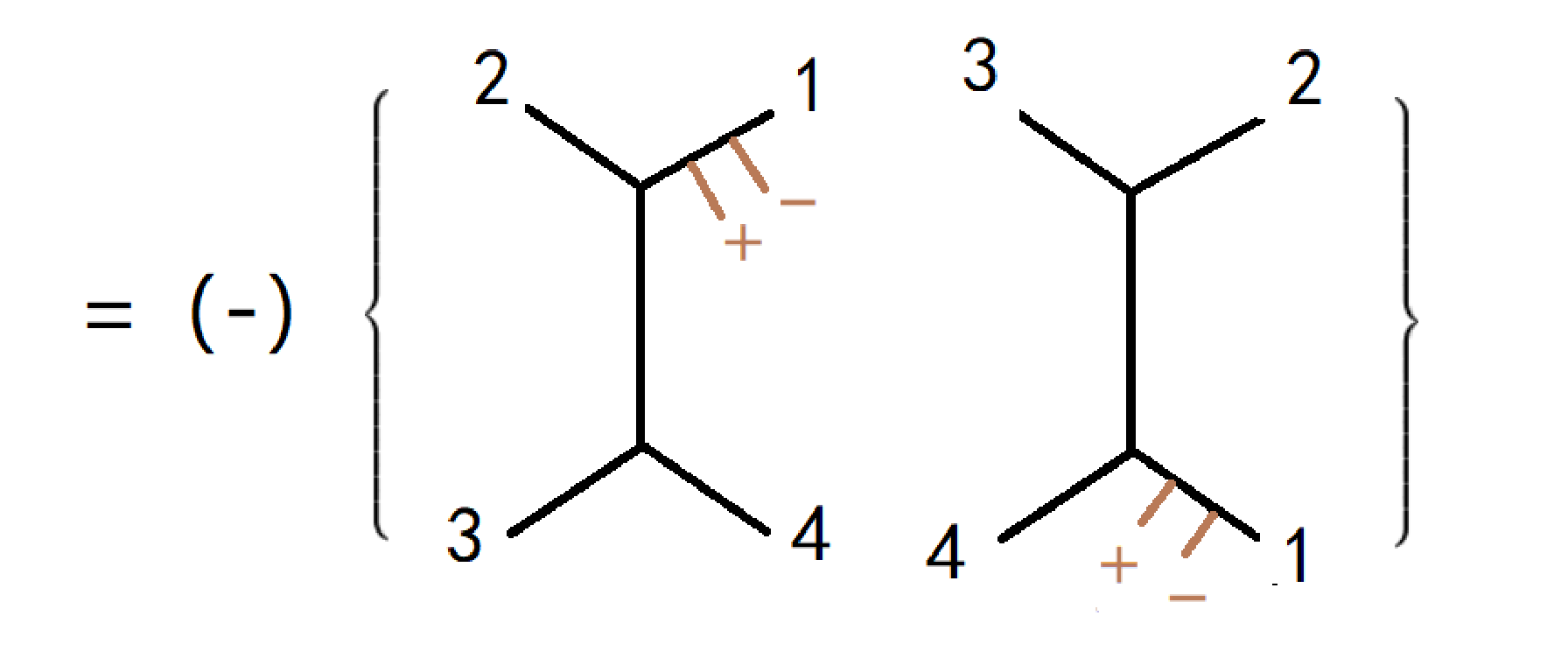}.
    \label{eq:1234bubble2}
\end{aligned}
\end{equation}
picking out all diagrams containing poles $s_{+-1}s_{-1}$.
Subtracting these singular contributions we get:
\begin{equation}
    \centering
    \includegraphics[scale=0.2]{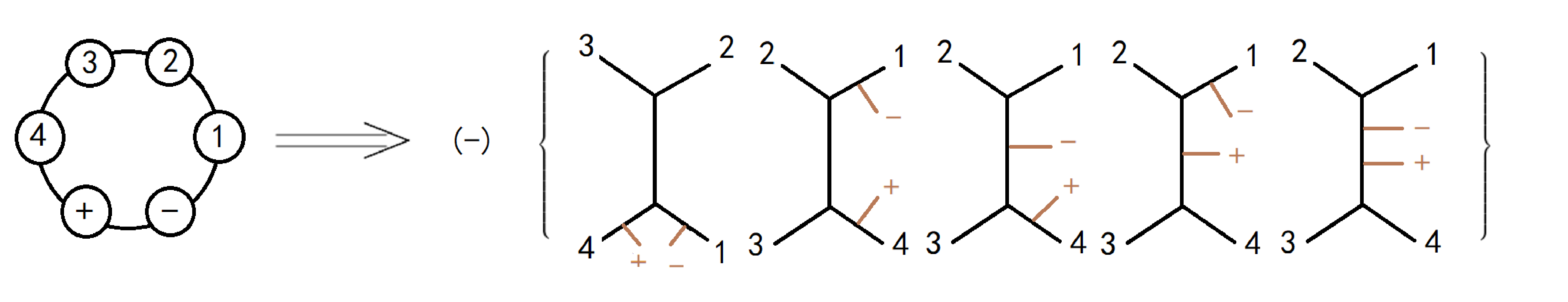}.
    \label{eq:1234+-}
\end{equation}
For  other three terms, we just need  to cyclically permutate
$(1,2,3,4)$ and obtain:
\begin{equation}
    \centering
    \includegraphics[scale=0.2]{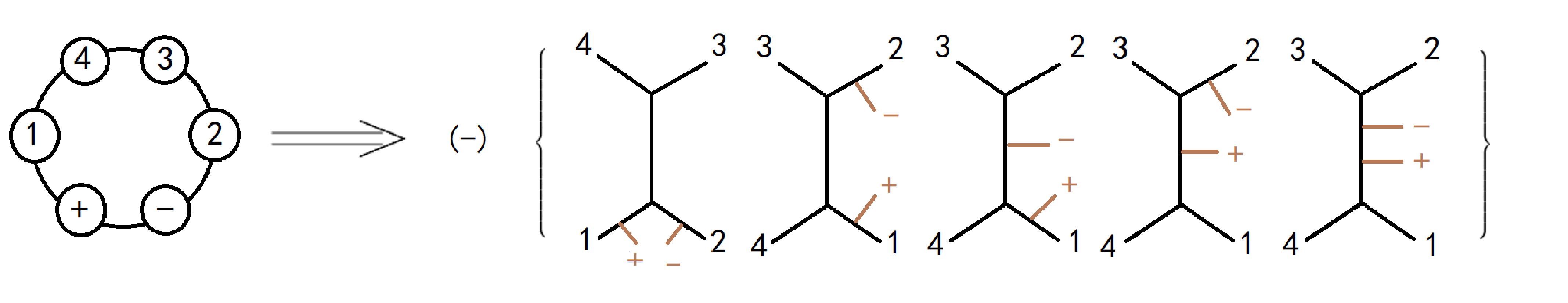};
    \label{eq:1+-234}
\end{equation}
\begin{equation}
    \centering
    \includegraphics[scale=0.2]{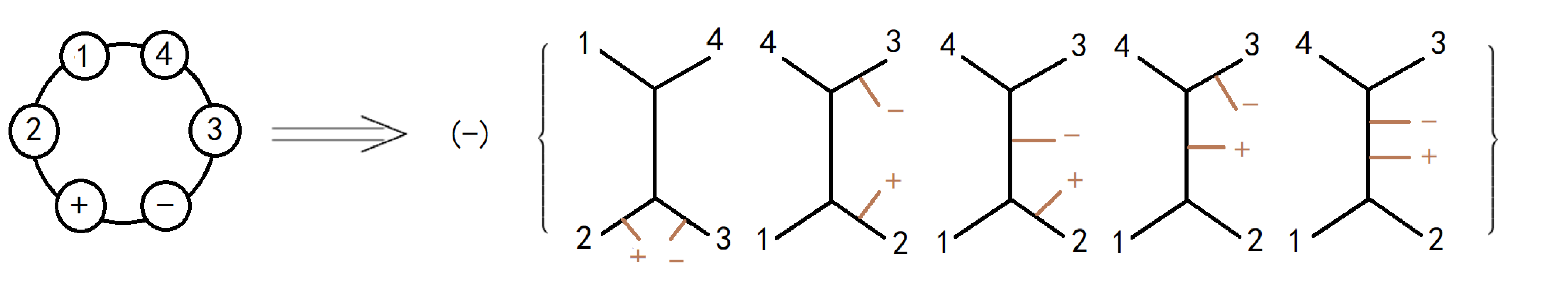};
    \label{eq:12+-34}
\end{equation}
\begin{equation}
    \centering
    \includegraphics[scale=0.2]{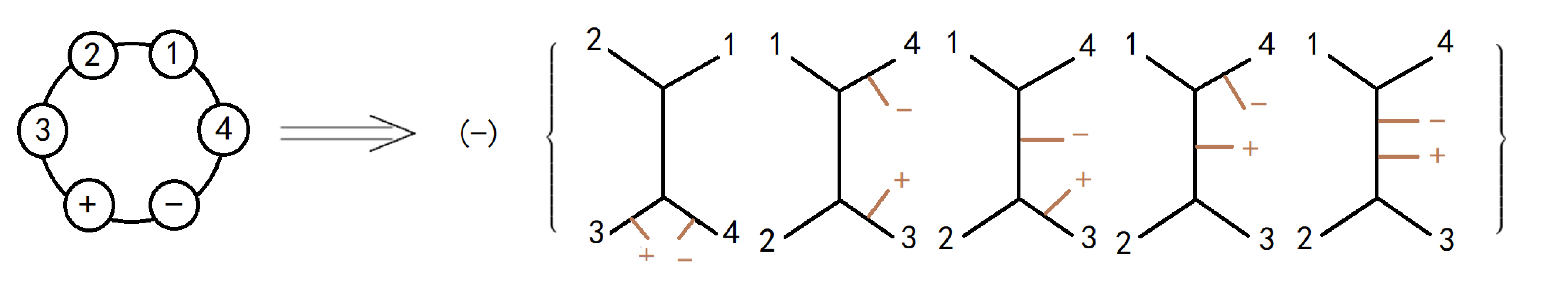}.
    \label{eq:123+-4}
\end{equation}
When we glue all $+,-$ legs as required by the forward limit, we
reach following one loop diagrams:
\begin{equation}
    \centering
    \includegraphics[scale=0.3]{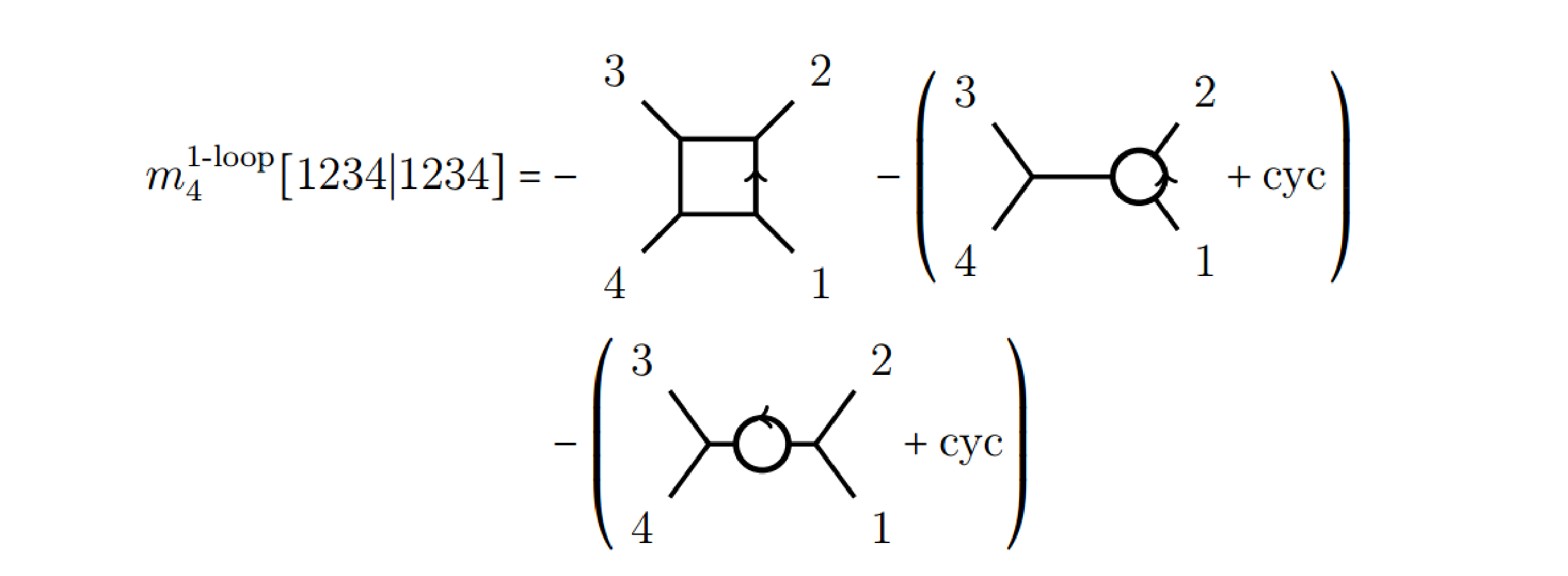}.
    \label{eq:new 4 point one loop}
\end{equation}
Thus we have presented four point example for \eqref{eq:formula}.
Compared with \eqref{eq:HY loop to tree}, the difference is: \bea
\Delta_4 m^{tree}_4[1234|1234]\equiv-\frac{1}{4}\int
\frac{d^Dl}{(2\pi)^D}\frac{1}{l^2}\sum^n_{a=1}\frac{1}{(l\cdot
k_a)^2}\times m^{tree}_4[1234|1234] ~~~~\label{eq:difference1} \eea
which follows from
\begin{equation}
    \includegraphics[scale=0.3]{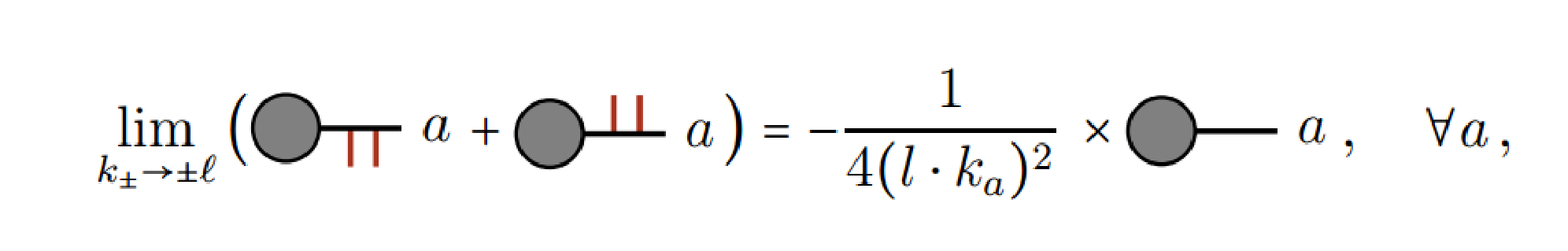}
    \label{eq:delta}
\end{equation}
Thus the difference between \eqref{eq:HY loop to tree} and
\eqref{eq:formula} is exactly those tree diagrams which contribute
to massless bubbles. Since its scale-free from, the integration of
\eref{eq:new 4 point one loop} is zero. Thus these two results are
in the same equivalent class. Compared with \eqref{eq:HY loop to
tree} with $n^2$ terms, for the special case $\pi=\rho$, we have
only $4n$ terms.  Furthermore, although the RHS of \eref{eq:delta}
is regular, each term at the LHS is singular, thus when we take the
forward limit.

\subsection{$\pi=\rho^T$}

Now we consider another special case, i.e., the two orderings are
opposite $\pi=\rho^T$, where $T$ means the reversing ordering of the
$\rho$. If it is for tree level amplitude, since
$m^{tree}_n[\pi|\rho]=(-)^nm^{tree}_n[\pi|\rho^T]$, we will have
$m^{tree}_n[\pi|\pi^T]=(-)^n m^{tree}_n[\pi|\pi]$. However, the
situation is different for one loop case. As shown in
\cite{He:2015yua}, $m^{1-loop}_4[1234|4321]$ contains only two
one-loop diagrams, while $m^{1-loop}_4[1234|1234]$ contains box,
triangle and bubble diagrams as given in \eref{eq:new 4 point one
loop}. The reason is that when we go from the ordering $\pi$ to the
ordering $\rho$, we need to flip legs at some cubic vertexes.
However, such flips are not allowed if the leg containing the loop
momentum.
\begin{figure}[ht]
    \centering
    \includegraphics[scale=0.4]{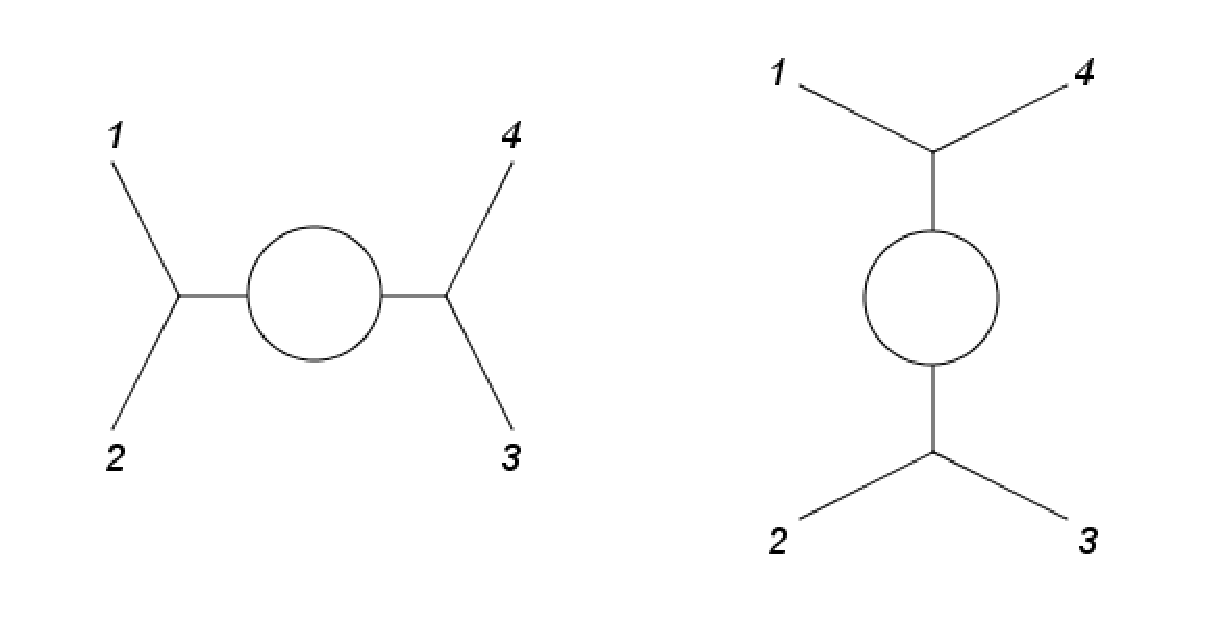}
    \caption{ $m^{1-loop}_4[1234|4321]$}
    \label{fig:1234and4321}
\end{figure}
Thus in this case, it is easy to see that all allowed one-loop
diagrams are massive bubble diagrams. The reason is following. If a
loop diagram is compatible with the $\pi$ ordering, then the $\pi$
ordering can be separated to $m$ groups $\{\Pi_1|
\Pi_2|...|\Pi_m\}$, where each group is attached to the loop at a
cubic vertex. When we do the flipping, we are allowed to flip only
inside each group, thus we can produce these orderings $\{{\cal
F}[\Pi_1]|{\cal F}[ \Pi_2]|...|{\cal F}[\Pi_m]\}$. Then if $m\geq
3$, it is impossible to go from $\pi$-ordering to $\pi^T$-ordering.
When $m=2$, it is easy to see that
\bea \pi\to \{\Pi_1|\Pi_2\}\to \{\Pi_1^T|\Pi_2^T\}=\{\Pi_2^T|\Pi_1^T\}=\pi^T.\eea
Furthermore, starting from the tree level diagrams, we can insert
the loop to each internal propagator. Collecting all of them, we get
the  $m^{1-loop}_n[\pi|\pi^T]$.

Now we discuss how to construct the one-loop CHY-integrand by above
understanding. We choose $\pi=(1,2,\cdots,n)$, $\rho=(n,\cdots,2,1)$
without loss of generality. Let us consider a typical diagram of the
amplitude $M^{1-loop}_n[1,2,\cdots,n|n,\cdots,2,1]$ drawing with the
clockwise ordering $(1,2,\cdots,n)$:
\begin{equation}
    \includegraphics[scale=0.5]{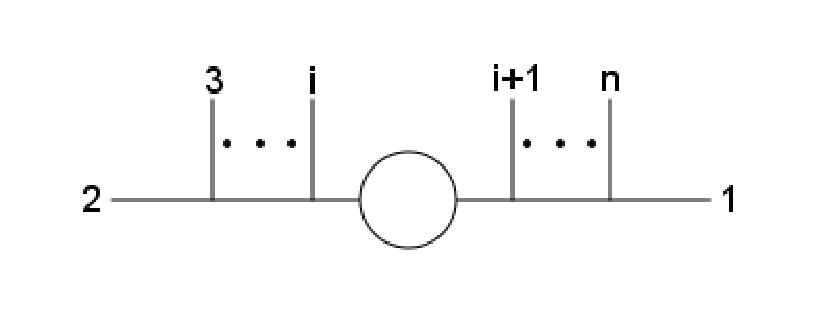}
    \label{eq:piorderoneloop}
\end{equation}
where $i$ can be any integer of $3\leq i\leq n-1$. After flipping,
from the diagram \eqref{eq:piorderoneloop} we can produce following
diagram with the $(n,\cdots,2,1)$-planar ordering
\begin{equation}
    \includegraphics[scale=0.5]{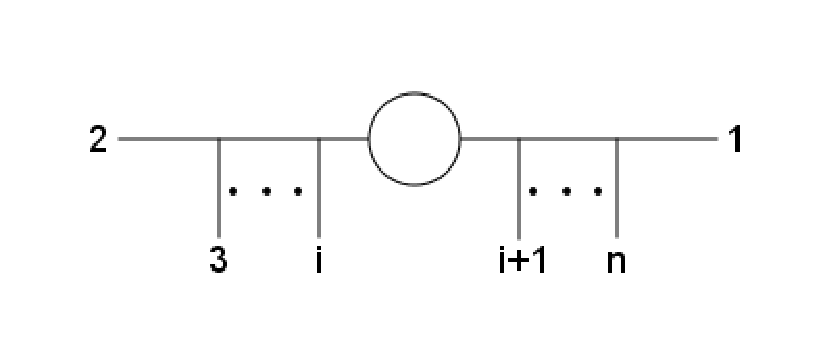}
    \label{eq:rhoorderoneloop}
\end{equation}
Cutting each loop propagator in \eqref{eq:piorderoneloop}, we get two
 $(n+2)$-point tree diagrams as following
\begin{equation}
    \includegraphics[scale=0.5]{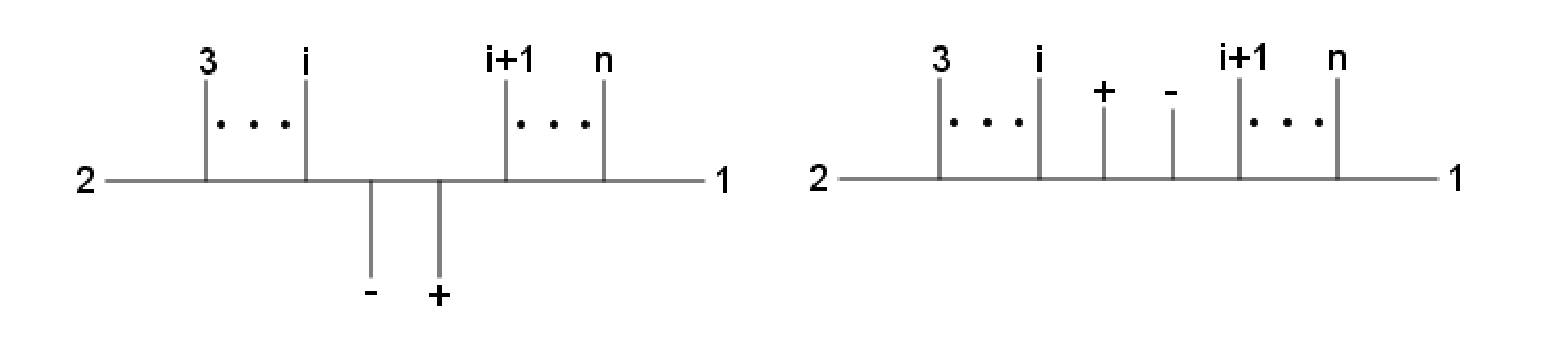}
    \label{eq:alphacut}
\end{equation}
while cutting the corresponding loop diagram
\eqref{eq:rhoorderoneloop} becomes
\begin{equation}
    \includegraphics[scale=0.5]{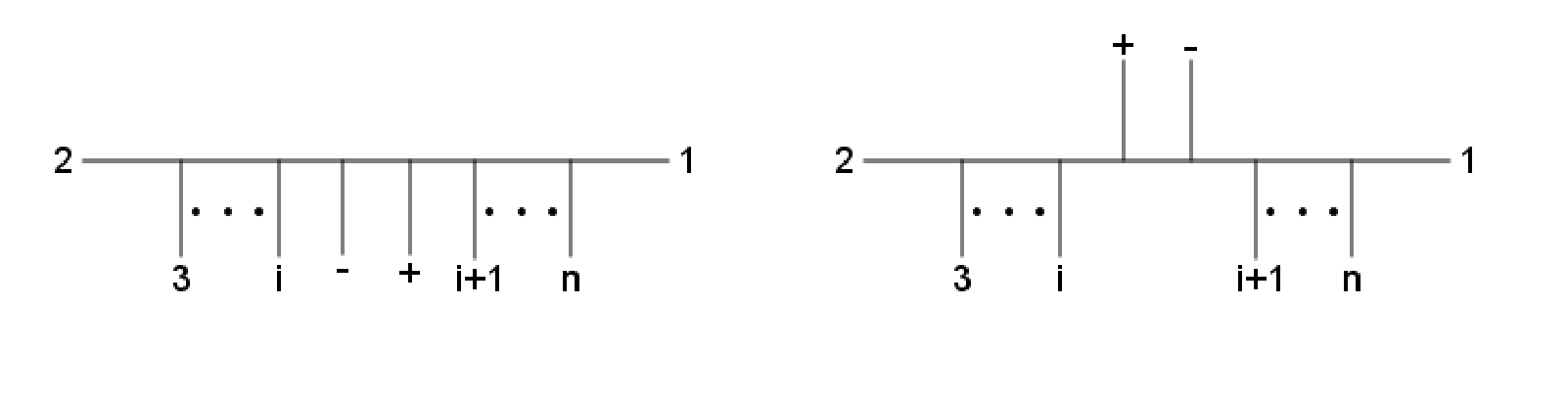}.
    \label{eq:betacut}
\end{equation}
An important point is that among these two diagrams
\eref{eq:alphacut} and \eref{eq:betacut}, there is a one-to-one
correspondence relation: the left one in \eqref{eq:alphacut}
corresponds to the left one  in \eqref{eq:betacut} and similarly the
right one to the right one. From the correspondence, we can see that
the left comes form the tree level CHY-integrand
\bea
PT(1,+,-,2,\cdots,n)PT(n,\cdots,i+1,+,-,i,\cdots,1),~~3\leq i\leq n-1.
\label{eq:legal}
\eea
It is easy to check that the CHY-integrand \eref{eq:legal} produces
terms containing tadpole $s_{+-}$, but not the massless bubbles.
Thus using the operator ${\cal P}^{+,1}_{-,2}$, we can pick up
tadpole contributions.

In above discussion, we have required $3\leq i\leq n-1$. Now we
consider the case $i=2$ and $i=n$. For these cases, the
corresponding CHY-integrands are given by
\bea
PT(1,+,-,2,\cdots,n)PT(n,\cdots,+,-,2,1)~~~~\label{eq:+-2}\\
PT(1,+,-,2,\cdots,n)PT(n,\cdots,2,1,+,-)~~~~\label{eq:1+-}
\eea
and their effective Feynman diagrams are following respectively
\begin{equation}
 \includegraphics[scale=0.3]{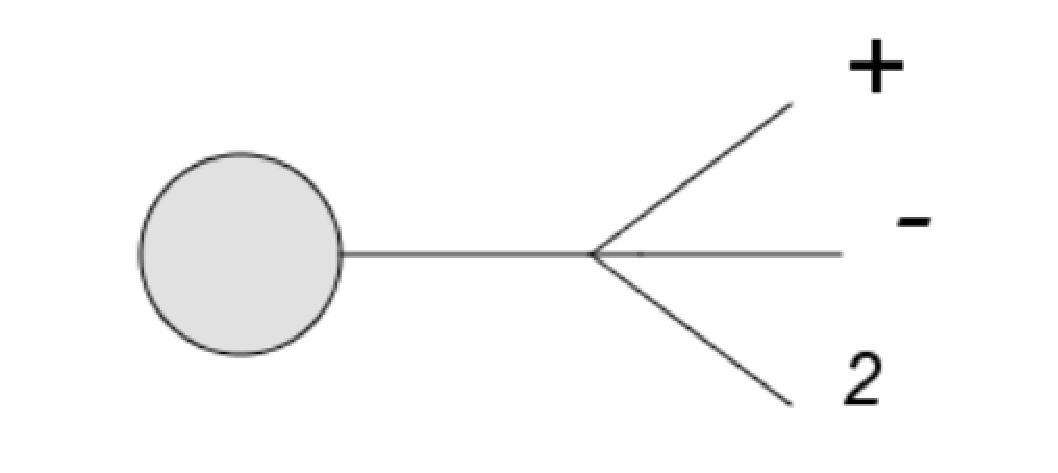} \label{iii-1}
\end{equation}
\begin{equation}
    \includegraphics[scale=0.3]{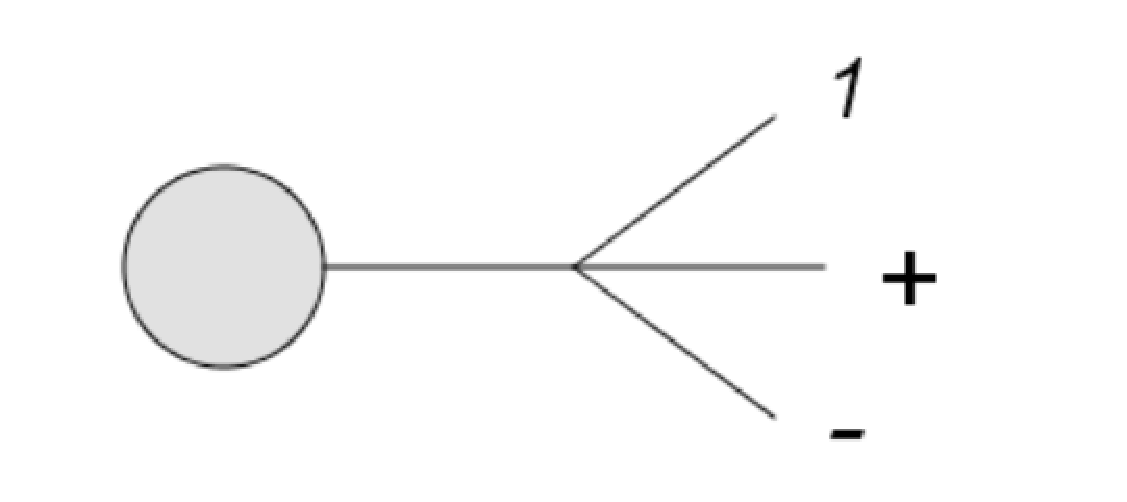} \label{iii-2}
\end{equation}
All these diagrams in \eref{iii-1} and    \eref{iii-2}  contain
either massless bubbles or tadpoles, thus we should not consider
these terms in the CHY-integrand. Now for a given insertion of $+,-$
legs in $\pi$-ordering, i.e., the fixed factor $PT(1,+,-,2,\cdots
,n)$, we have considered all possible insertion in the
$\pi^T$-ordering, except the following one $PT(n,\cdots ,2,+,-,1)$.
It is easy to see that
\begin{equation}
    PT(1,+,-,2,\cdots ,n)PT(n,\cdots ,2,+,-,1)={\cal P}^{+,2}_{-,1}PT(1,+,-,2,\cdots ,n)PT(n,\cdots ,2,-,+,1)
    \label{eq:inverse1+-2}
\end{equation}
which means all Feynman diagrams comings from \eqref{eq:inverse1+-2}
have pole structure $s_{+-}$. In other words, it gives only tadpole
diagrams, so we should drop this term.

Above discussion tells us what will happen when one insert $+,-$
legs between $\{1,2\}$ in the $\pi$-ordering. Similar result can be
obtained for the insertion between another pair $\{i,(i+1)\}$. When
we sum up all these pairs together, we arrive following proposal for
one loop CHY-integrand of $m^{1-loop}[\pi|\pi^T]$
\begin{equation}
\begin{aligned}
    &m^{1-loop}_n[1,2,\cdots, n|n,\cdots,2,1]\\=&\int\frac{d^Dl}{(2\pi)^D}\frac{1}{l^2}\lim_{k_{\pm}\mapsto\pm l}\int d\Omega_{CHY}\sum_{i,j\neq i,i\pm 1}(1-{\cal P}^{+,i}_{-,i+1})PT(\cdots ,i,+,-,i+1,\cdots)PT(\cdots j+1,+,-,j,\cdots).
\label{eq:inverse}
\end{aligned}
\end{equation}
where the factor $(1-{\cal P}^{+,i}_{-,i+1})$ removes the tadpole contribution effectively.

Now we present an  example, i.e., $\pi=(1234)$ and $\rho=(4321)$. Before taking
 the forward limit, the RHS of \eqref{eq:inverse} gives following four terms
\begin{equation}
    \begin{aligned}
        &(1-{\cal P}^{+,1}_{-,2})PT(1,+,-,2,3,4)PT(4,+,-,3,2,1)\\
        +&(1-{\cal P}^{+,2}_{-,3})PT(1,2,+,-,3,4)PT(4,3,2,1,+,-)\\
        +&(1-{\cal P}^{+,3}_{-,4})PT(1,2,3,+,-,4)PT(4,3,2,+,-,1)\\
        +&(1-{\cal P}^{+,4}_{-,1})PT(1,2,3,4,+,-)PT(4,3,+,-,2,1).\\
    \end{aligned}
\end{equation}
Each term gives one diagram in following result
\begin{equation}
    \includegraphics[scale=0.2]{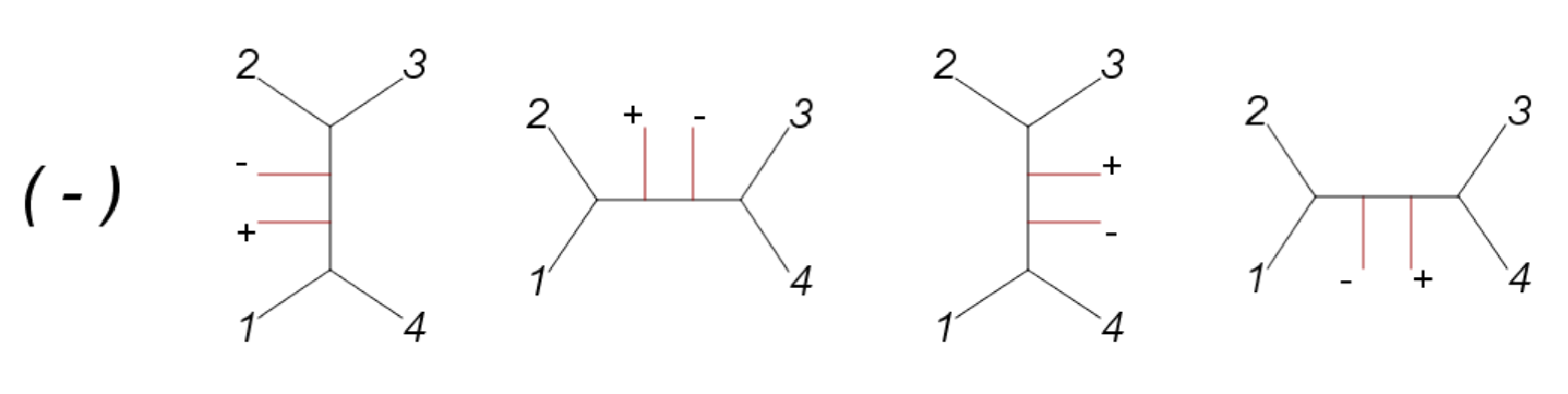}
    \label{eq:reversed1234}
\end{equation}
Since there is no any singularity left, one can take the forward limit and obtain
\begin{equation}
    \includegraphics[scale=0.33]{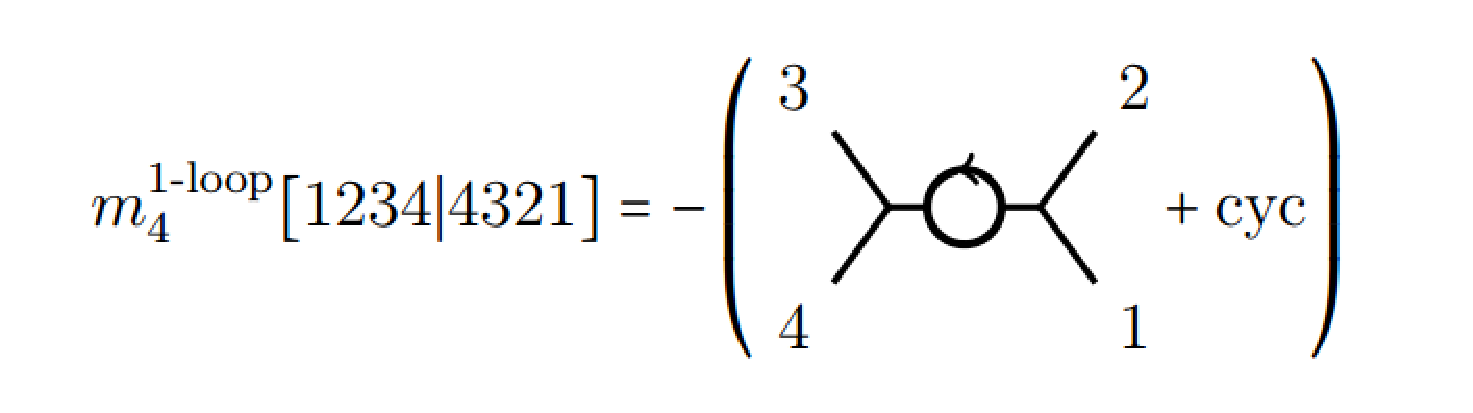}.
    \label{eq:reslut4321}
\end{equation}
Compared with the proposal \eqref{eq:HY loop to tree}, again we do
not have following massless bubble terms
\bea
\Delta_4 m^{tree}_4[1234|4321]\equiv-\frac{1}{4}\int \frac{d^Dl}{(2\pi)^D}\frac{1}{l^2}\sum^n_{a=1}\frac{1}{(l\cdot k_a)^2}\times m^{tree}_4[1234|4321]
~~~~\label{eq:difference2}
\eea

\subsection{For general orderings $\pi$ and $\rho$}

Having dealt with two special cases, now we move to  general cases,
i.e., with arbitrary orderings of $\pi$ and $\rho$. The first
important observation is that $m^{1-loop}_n[\pi|\rho]\neq 0$ if and
only if $m^{tree}_n[\pi|\rho]\neq 0$. The argument is follows. If
$m^{tree}_n[\pi|\rho]\neq 0$, we can obtain one-loop diagrams
respecting  both $\pi$- and $\rho$-planar orderings by adding an
bubble on any internal propagator. On the other hand, given any
one-loop diagram in $m^{1-loop}_n[\pi|\rho]$, we can construct the
corresponding tree diagrams respecting both $\pi$- and $\rho$-planar
orderings. The procedure is follows. At the first step, we shrink
the loop to a point, so we get an effective vertex. Then we can
expand the effective vertex to any tree diagram with only cubic
vertexes. Above argument is illustrated  in the Figure \ref{fig:two
steps}.

\begin{figure}[ht]
    \centering
    \includegraphics[scale=0.35]{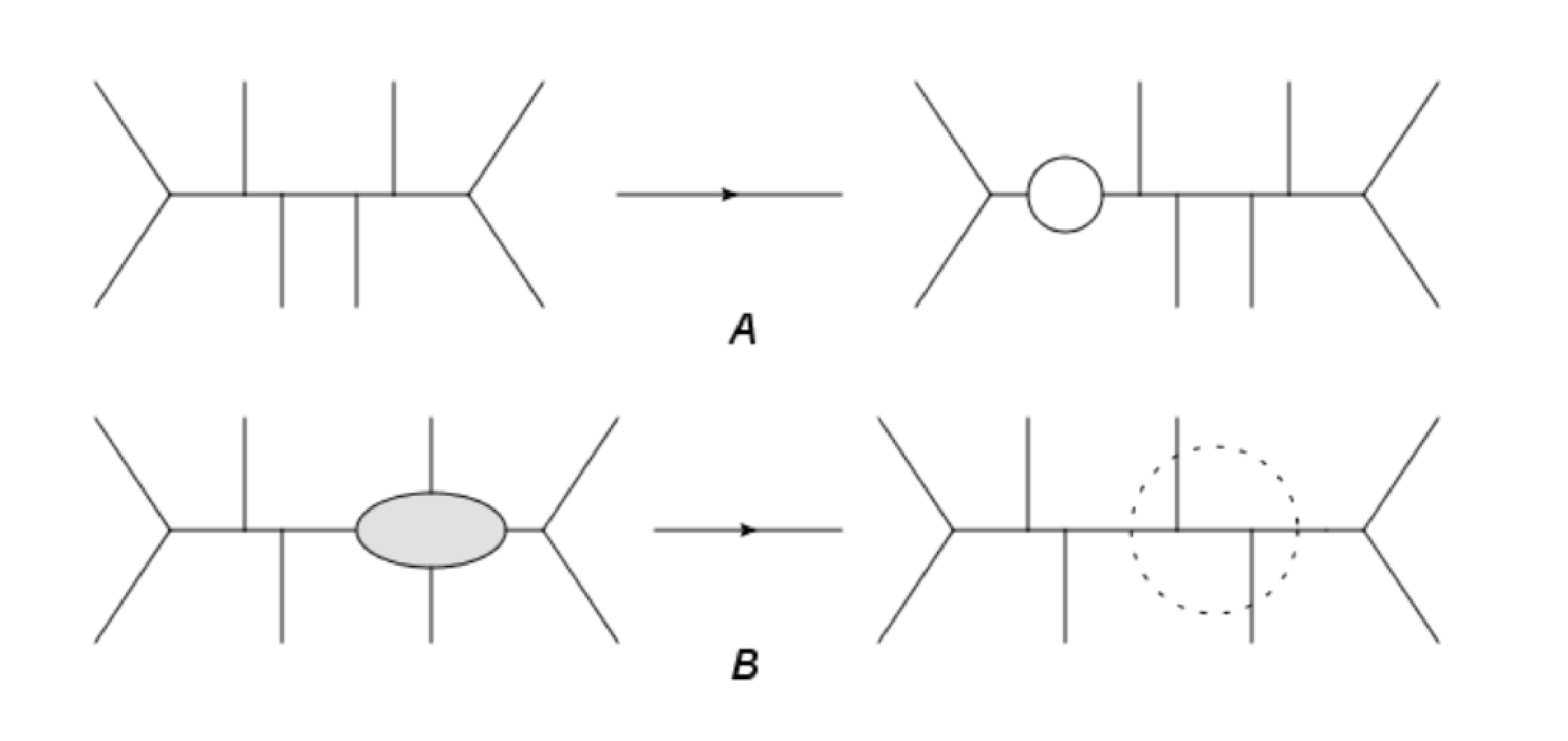}
    \caption{(A) The way constructing one-loop diagram from tree. (B)
    The way constructing tree diagram from one-loop diagram}
    \label{fig:two steps}
\end{figure}

With above observation, now we show how to construct the one-loop
CHY-integrand from the corresponding tree-level effective Feynman
diagram. For clarity, we will use the example
$\pi=(1,2,3,4,5,6,7,8)$ and $\rho=(1,2,8,4,6,5,7,3)$ to demonstrate
our construction.

\begin{figure}[ht]
    \centering
    \includegraphics[scale=0.35]{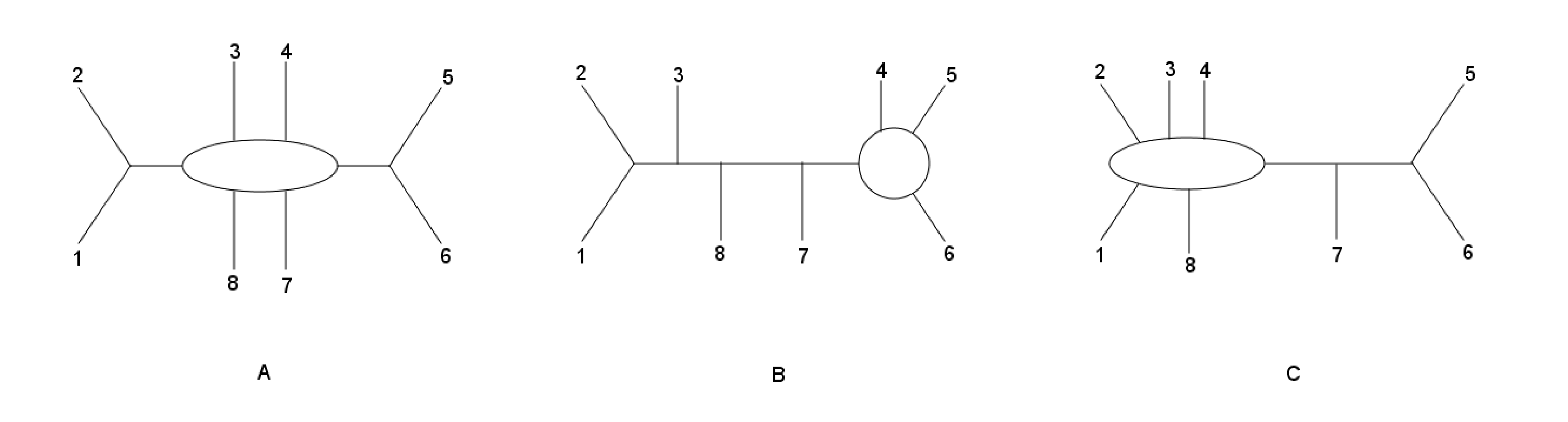}
    \caption{Some one loop Feynman diagrams cover more than one effective vertexes}
    \label{fig:illegueoneloop}
\end{figure}

At the first step, we insert a loop to the effective tree level
Feynman diagram. A key point is that after the insertion of the
loop, no any effective propagator should be replaced. Let us see
what will happen if an effective propagator has been replace. For
example, we have drawn  three graphs in the Figure
\ref{fig:illegueoneloop} according to the planar ordering $\pi$ of
above example, where the loop in the graphs $A,B$ has replaced the
effective propagator $s_{1238}, s_{56}$ respectively while  the loop
in the graphs $C$ has replaced two effective propagators $s_{1238},
s_{12}$. It is easy to see that by flips, we can not transform the
planar ordering $\pi$ to the ordering $\rho$ (remembering that we
can not flip lines containing loop momentum). The reason is simple.
As we have shown before, two effective vertexes connecting by a
given effective propagator must have opposite ordering, i.e., one is
the $\cal{SO}$-type and another one, $\cal{RO}$-type.
 Thus to transform the $\pi$-ordering to the
$\rho$-ordering, we need to flip at the vertex having reversing
ordering. However, if such an effective propagator has been replaced
by the loop, the needed flips do not exist, thus at the loop level,
we can not transform $\pi$-ordering to $\rho$-ordering.

Above clarification shows that there are two types of insertions of
the loop. The first type is to inserted on the effective propagators
as well as on the external lines\footnote{Here we have included the
massless bubbles when considering the construction of
CHY-integrands. Later we will move these singularities away.  }
    \begin{equation}
        \includegraphics[scale=0.35]{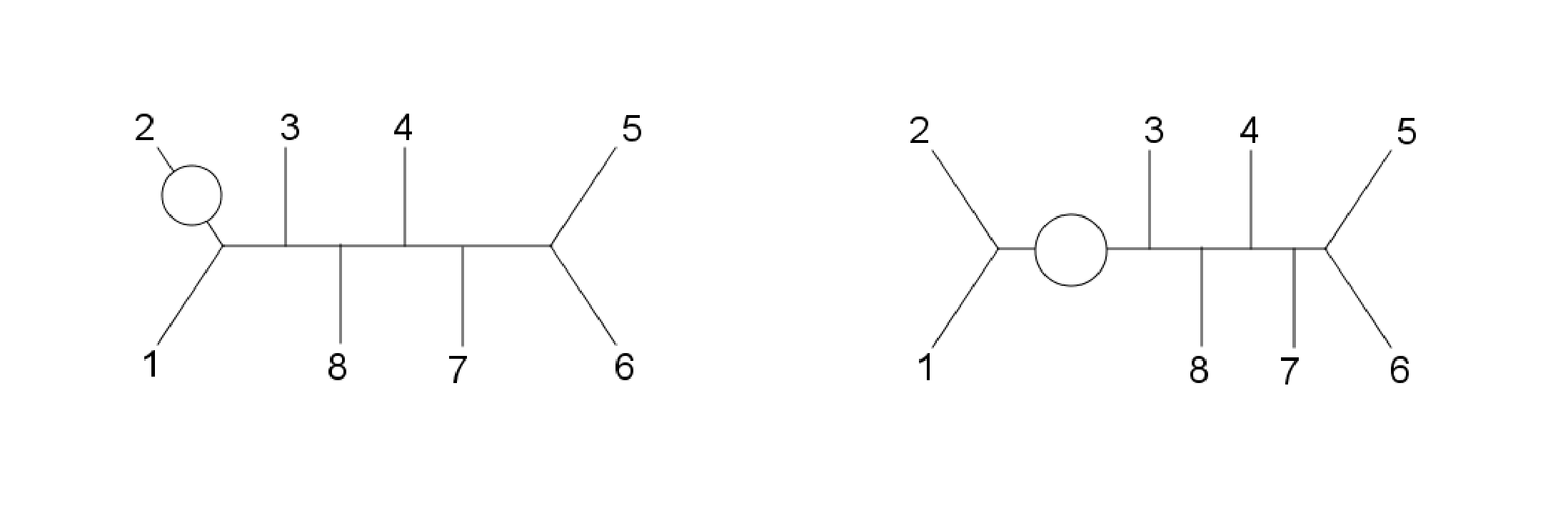}
        \label{eq:cover0vertex}
    \end{equation}
The second type is rest cases, where the loop is not inserted either on the effective propagators or on the external lines as shown in following:
\begin{equation}
    \includegraphics[scale=0.35]{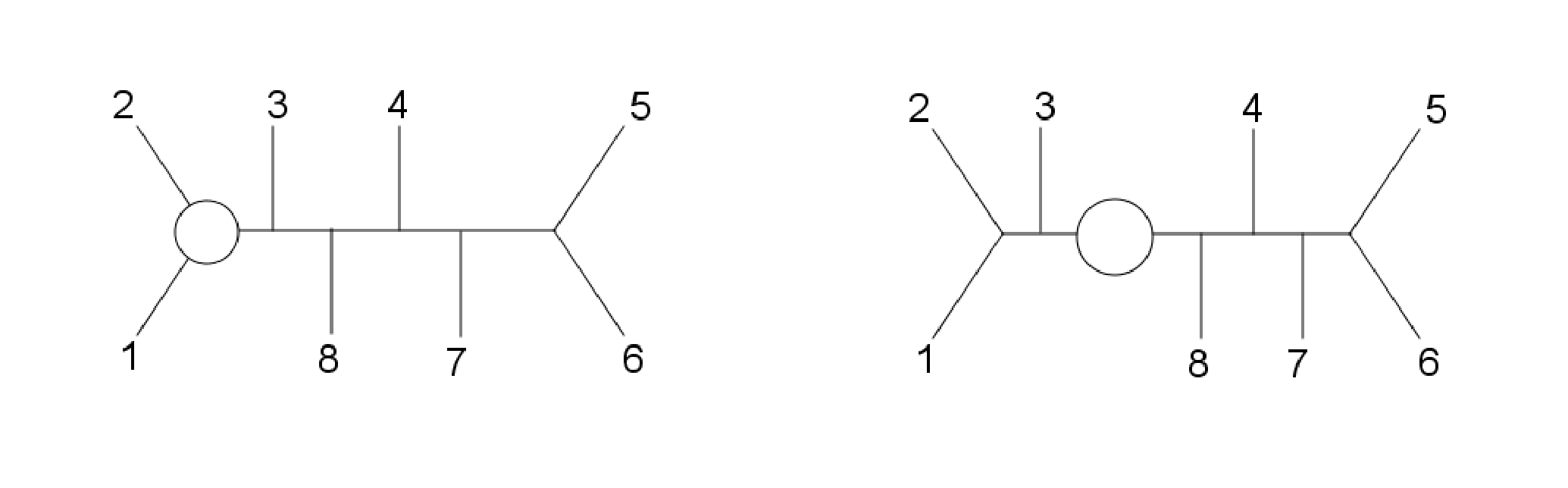}
    \label{eq:cover1vertex}
\end{equation}

Having classified the type of loops, we can attribute the inserted
loops to the corresponding effective vertexes. For the second type
of loops, it must belong to a particular effective vertex. For the
first type of loops, there are two cases. If it is inserted on the
effective propagator, it can be attribute to each of effective
vertexes at the two ends of the propagator. If it is inserted on the
external legs, it must belong to a particular effective vertex. With
such an assignment,  all loop Feynman diagrams connecting to an
effective vertex $V_i$ can be collected together and denoted as $\WH
m^{1-loop}_{{\cal SO}/{\cal RO}}[V_i]$ where ${\cal SO},{\cal RO}$
mean the ${\cal SO}, {\cal RO}$ type vertexes respectively. For
instance, $\WH m^{1-loop}_{{\cal SO}}[V_1]$ and $\WH
m^{1-loop}_{{\cal RO}}[V_2]$ in our example are shown as follows.
\begin{equation}
    \includegraphics[scale=0.33]{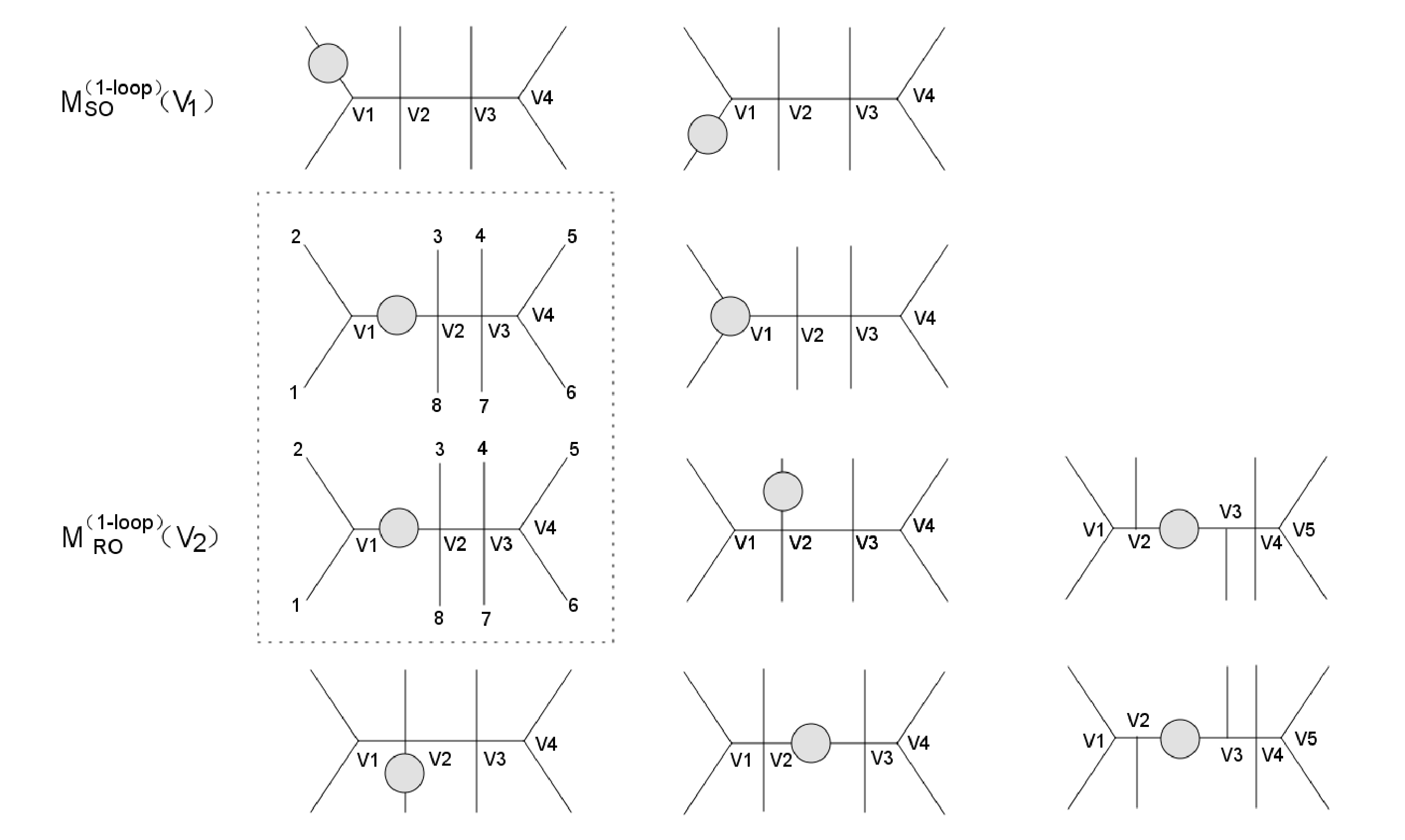}
    \label{eq:MV}
\end{equation}
With such a grouping, we can write
\begin{equation}
    \WH m^{1-loop}_n[\pi|\rho]=\bigcup_{V_i}\WH m^{1-loop}_{{\cal SO}/{\cal RO}}[V_i]
    \label{eq:merge}
\end{equation}
where we use union instead of sum because as we have mentioned
before, the loop inserted on the effective propagator has been
assigned to two effective vertexes as shown in   \eqref{eq:MV} by
the dashed box. Now the problem is reduced to find  out the
CHY-integrands for each $\WH m^{1-loop}_{SO/RO}[V_i]$'s.

For an arbitrary effective vertex $V_i$ as
\begin{equation}
    \includegraphics[scale=0.45]{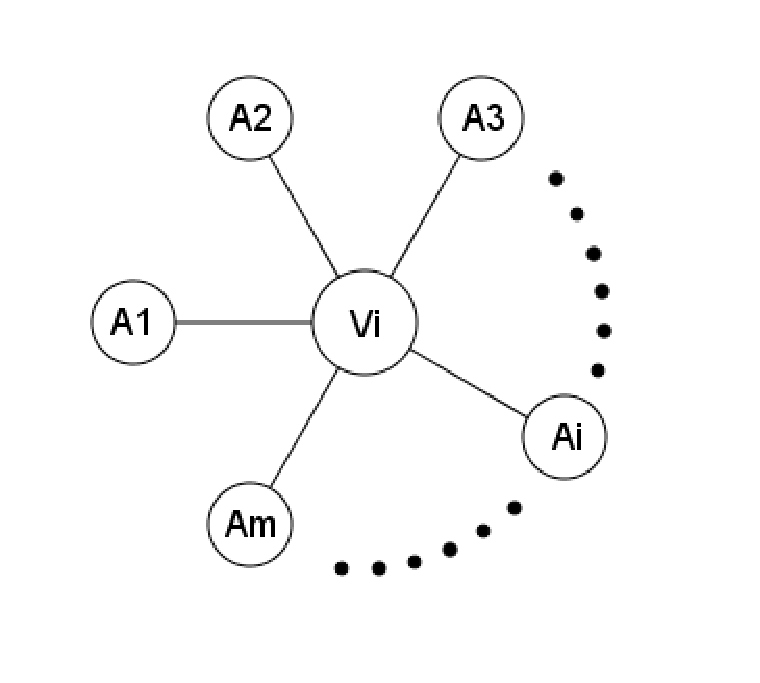}
    \label{eq:effectVi}
\end{equation}
if it is the ${\cal SO}$-type, the  $\pi$- and $\rho$-planar
ordering splitting must be the form
\begin{equation}
\begin{aligned}
    &\pi-split=(A_1)(A_2)\cdots(A_m)\\
    &\rho-split=(A'_1)(A'_2)\cdots(A'_m)
    \label{eq:V-split-same}
\end{aligned}
\end{equation}
where $A'_i$ is a permutation\footnote{ $A'_i$ can not be an
arbitrary permutation of $A_i$, since they must be connected by some
flips at the cubic vertexes. } of $A_i$. If it is the ${\cal
RO}$-type, the splitting must be
\begin{equation}
\begin{aligned}
    &\pi-split=(A_1)(A_2)\cdots(A_m)\\
    &\rho'-split=(A'_m)\cdots(A'_2)(A'_1)
    \label{eq:V-split-reversed}
\end{aligned}
\end{equation}
When all $A_i$ in \eqref{eq:V-split-same} and
\eqref{eq:V-split-reversed} contain only one external leg, it is
just reduced to the two special cases $\WH m^{1-loop}_n[1,2,\cdots
,n|1,2,\cdots, n]$ or $\WH m^{1-loop}_n[1,2,\cdots, n|n,\cdots,2,1]$
studied in previous subsections. Now it is clear how to extend the
construction to general cases: we treat each group $A_i$ as a single
element and insert $(+,-)$  between each neighboring group  in both
orderings at same time. More explicitly,  in the case
\eqref{eq:V-split-same}, we can construct formally as follows:
\begin{equation}
\begin{aligned}
   &\WH m^{1-loop}_{SO}[V]=\sum_{i}(1-{\cal P}^{+,\cdot}_{-,\cdot})PT(\cdots(A_i)+-(A_{i+1})\cdots)PT(\cdots(A'_i)+-(A'_{i+1})\cdots)
   \label{eq:sequential vertex}
\end{aligned}
\end{equation}
where $\cdot$ in $\cal{P}$ denote the right-most particle next to
$+l$ and left-most particle next to $-l$. While in the case
\eqref{eq:V-split-reversed}, we have formally
\begin{equation}
    \begin{aligned}
        \WH m^{1-loop}_{RO}[V]\subseteq \sum_{i=1}^{m}\{(1-{\cal P}^{+,\cdot}_{-,\cdot})\sum_{j\neq i}PT(\cdots(A_i),+,-,(A_{i+1}),\cdots)PT(\cdots(A'_{j+1}),+,-,(A'_j)\cdots)\}.
    \end{aligned}
    \label{eq:inverse vertex}
\end{equation}
In \eref{eq:sequential vertex} and \eref{eq:inverse vertex}, the combination $(1-{\cal P}^{+,\cdot}_{-,\cdot})$ has been used to eliminate the tadpole contributions. Furthermore, in \eref{eq:inverse vertex} the summation region is larger than the one in \eref{eq:inverse} where $j\neq i, i\pm 1$. The reason is follows. An external line of an effective vertex could be the effective propagator, thus the inserted loop on this line should be kept. Because the same reason, in \eref{eq:sequential vertex} we have not removed the external bubbles.

\begin{figure}[ht]
    \centering
    \includegraphics[scale=0.45]{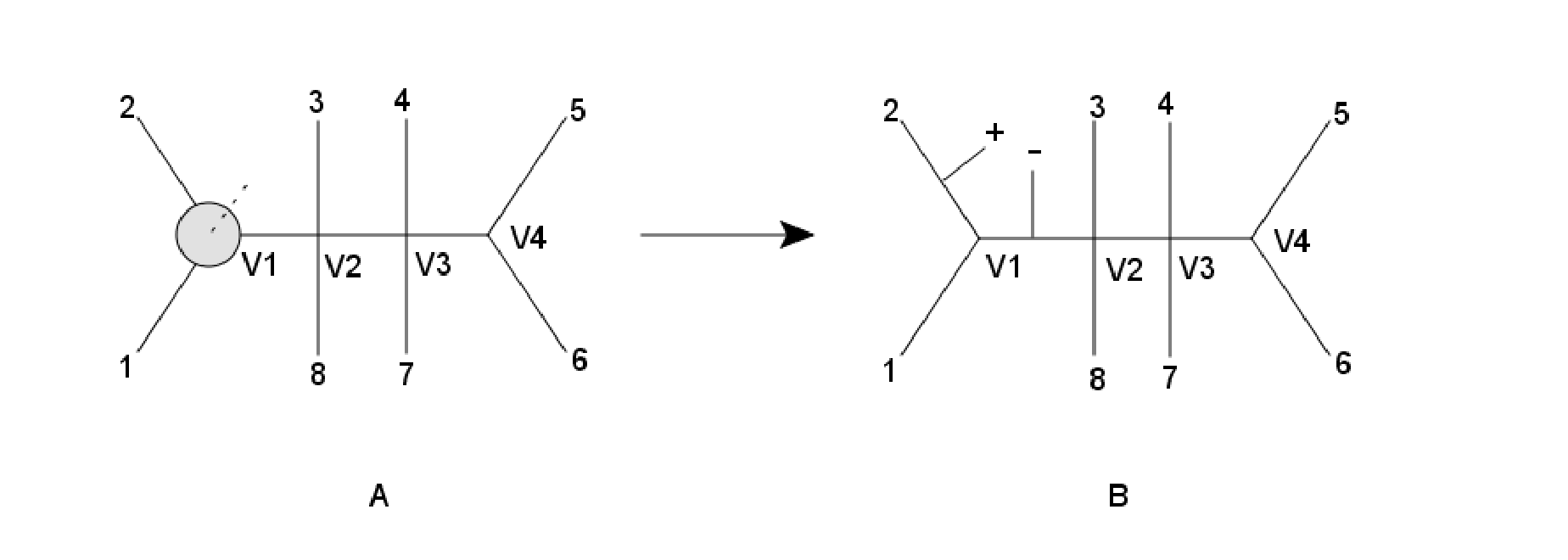}
    \caption{The redundancy of two neighbouring effective vertex}
    \label{fig:redundant}
\end{figure}

Another important point in  \eref{eq:inverse vertex} is that  we use the notation $\subseteq$ instead of the one $=$, because the right hand side of  \eqref{eq:inverse vertex} may contain extra diagrams which do not belong to the left hand side. In our example, when we analyze the effective vertex $V_2$, it will contain
\bea
(1-{\cal P}^{+,2}_{-,3})PT[(1,2),+,-,(3),(4,5,6,7),(8)]PT[(1,2),+,-,(8),(4,6,5,7),(3)].~~~~\label{redundant}\eea
This term contains a $10$-point tree diagram as shown in graph B of Figure \ref{fig:redundant}, which contributes to $\WH m^{1-loop}_{{\cal SO}}[V_1]$ since it is obtained by cutting an edge of one loop diagram (graph A of Figure \ref{fig:redundant}).
Furthermore, there are some terms appearing in both \eref{eq:sequential vertex} and \eref{eq:inverse vertex}. For example, \eqref{redundant} will appear in both  $\WH m^{1-loop}_{{\cal SO}}[V_1]$ and $\WH m^{1-loop}_{{\cal RO}}[V_2]$. When we sum contributions from all effective vertexes, these repeated terms should be included only once.


Finally, we move to the problem of how to dislodge massless bubble part. The tree diagrams after cutting a massless bubble are shown in Figure \ref{fig:tadpole and bubble}, from which we see that   all $(n+2)$-point tree diagrams obtained by cutting massless bubbles are included in following type of terms:
\begin{equation}
    \begin{aligned}      &PT(\cdots,i,+,-,  \cdots)PT(\cdots,i,+,-,\cdots)\ \ 
        or\  &PT(\cdots,+,-,i,\cdots)PT(\cdots,+,-,i,\cdots)
    \end{aligned}
    \label{eq:externalbubble}
\end{equation}
Now we can apply the method of picking out pole to eliminate them. There are two cases. In the first case, all Feynman diagrams produced by a  CHY-integrand will  contain the $s_{i+-}$ or $s_{+-i}$ poles. For these CHY-integrands, we just remove them from the sum. In the second case, $s_{i+-}$ pole will appear only on some Feynman diagrams (but not all). Thus we need to  multiply two cross ratio factor ${\cal P}{\cal P}$ to pick up these terms as we did in \eqref{eq:formula}.

Now we need to understand when all Feynman diagrams produced by a  CHY-integrand will  contain the $s_{i+-}$ pole. It is easy to see their effective Feynman diagrams  will be these in \eref{iii-1} and \eref{iii-2}, thus  if using the method given in \cite{Cachazo:2013iea}, will be following diagrams
\begin{equation}
    \includegraphics[scale=0.5]{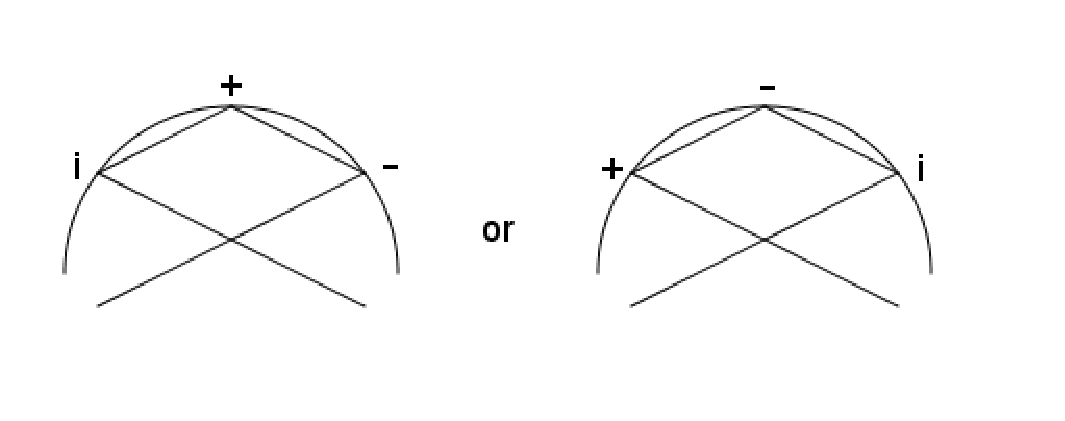}.
    \label{fig:i+-CHY}
\end{equation}
This structure of  the left hand graph in \eref{fig:i+-CHY}  means that in one PT-factor, the ordering is $(...,a,i,+,-,b,...)$, while in another PT-factor, the ordering is $(..., b,..., i,+,-, ...,a,...)$. Similar conclusion holds for the right hand graph. In our example with $\pi=(3,4,5,6,7,8,1,2)$, $\rho=(3,1,2,8,4,6,5,7)$. following terms
\begin{equation}
    \begin{aligned}
        &PT(1,2,\underline{+,-,3},4,5,6,7,8,9)PT(1,2,8,4,6,5,7,\underline{+,-,3})\\
        and\ &PT(1,2,\underline{3,+,-},4,5,6,7,8,9)PT(1,2,8,4,6,5,7,\underline{3,+,-}).
    \end{aligned}
\end{equation}
should be removed by our criterion \eref{fig:i+-CHY} for the pole $s_{3+-}$.

From the criterion \eqref{fig:i+-CHY}, we see directly that all external legs attached to $\cal{RO}$-type effective vertexes should be removed while all those attached to $\cal{SO}$-type vertex should be kept. In addition, if we attribute those diagrams, where the loop is inserted on an effective propagator, only to
the $\cal{SO}$-type vertex attached to the propagator,  we can rewrite \eqref{eq:inverse vertex} as:
\begin{equation}
    \begin{aligned}
        \WH m^{1-loop}_{{\cal RO}}[V]= \sum_{i=1}^{m}\{(1-{\cal P}^{+,\cdot}_{-,\cdot})\sum_{j\neq i,i\pm 1}PT(\cdots(A_i),+,-,(A_{i+1}),\cdots)PT(\cdots(A'_{j+1}),+,-,(A'_j)\cdots)\}.
    \end{aligned}
    \label{eq:inverse vertex01}
\end{equation}
Now we can use the notation $=$ since in \eref{eq:inverse vertex01} all mentioned problems have been fixed.
Moreover, there are no repeated terms appearing in both \eqref{eq:sequential vertex} and \eqref{eq:inverse vertex01}. Therefore, \eqref{eq:merge} can be rewrote as
\begin{equation}
    \WH m^{1-loop}_n[\pi|\rho]=\sum_{V_i}\WH m^{1-loop}_{{\cal SO}/{\cal RO}}[V_i]
    \label{eq:merge01}
\end{equation}

Up to now, we have discussed all technical points of our construction. Now we summarize our algorithm for
general $m^{1-loop}_n[\pi|\rho]$:
\begin{itemize}

\item (1) At the first step, we draw the $n$-point tree-level effective Feynman diagram of $m^{tree}_n[\pi|\rho]$.

\item (2) For each effective vertex, we determine its relative ordering, i.e., either $\cal{SO}$-type or $\cal{RO}$-type.

\item (3) For each effective vertex, we apply formula \eqref{eq:sequential vertex} or
\eqref{eq:inverse vertex01} to get CHY-integrands.

\item (4) Now we add those terms coming from different effective vertexes together as given in \eqref{eq:merge01}.

\item (5) For terms from previous step having the pattern in \eqref{eq:externalbubble}, we need to remove the massless bubble contributions using the cross ratio factor as did \eqref{eq:formula}.

\item (6) After finishing above steps, we finally get the CHY-integrand we are looking for.

\end{itemize}

\subsection{Example}

In this subsection, we present two examples to demonstrate our algorithm.

\subsubsection{The first example}
The first example is $m^{1-loop}_8[12345678|12846573]$ with its  effective tree Feynman diagram  shown in \eqref{eq:effect12846573} where $V_1, V_3$ are $\cal{SO}$-type and $V_2, V_4$, $\cal{RO}$-type.
For $V_1$, we have
\begin{equation}
    \begin{aligned}
        &(1-{\cal{P}}^{+,1}_{-,2})PT[(1)+-(2)(345678)]PT[(1)+-(2)(846573)]\\
        &(1-{\cal{P}}^{+,2}_{-,3})PT[(1)(2)+-(345678)]PT[(1)(2)+-(846573)]\\
        &(1-{\cal{P}}^{+,8}_{-,1})PT[(1)(2)(345678)+-]PT[(1)(2)(846573)+-].\\
    \end{aligned}
    \label{eq1:V_1}
\end{equation}
For $V_2$, we have:
\begin{equation}
    \begin{aligned}
        &(1-{\cal{P}}^{+,2}_{-,3})PT[(12)+-(3)(4567)(8)]PT[(12)(8)+-(4657)(3)]\\
        &(1-{\cal{P}}^{+,3}_{-,4})PT[(12)(3)+-(4567)(8)]PT[(12)+-(8)(4657)(3)]\\
        &(1-{\cal{P}}^{+,7}_{-,8})PT[(12)(3)(4567)+-(8)]PT[(12)(8)(4657)(3)+-]\\
        &(1-{\cal{P}}^{+,8}_{-,1})PT[(12)(3)(4567)(8)+-]PT[(12)(8)(4657)+-(3)]
    \end{aligned}
    \label{eq:V_2}
\end{equation}
For $V_3$, we have:
\begin{equation}
    \begin{aligned}
       &(1-{\cal{P}}^{+,3}_{-,4})PT[(8123)+-(4)(56)(7)]PT[(3128)+-(4)(65)(7)]\\
       &(1-{\cal{P}}^{+,4}_{-,5})PT[(8123)(4)+-(56)(7)]PT[(3128)(4)+-(65)(7)]\\
       &(1-{\cal{P}}^{+,6}_{-,7})PT[(8123)(4)(56)+-(7)]PT[(3128)(4)(65)+-(7)]\\
       &(1-{\cal{P}}^{+,7}_{-,8})PT[(8123)(4)(56)(7)+-]PT[(3128)(4)(65)(7)+-]\\
    \end{aligned}
    \label{eq1:V_3}
\end{equation}
For $V_4$, we have no contribution. Assembling above, we have $11$ terms in total. By applying picking operator $\cal P$ to terms containing massless bubbles, we obtain:
\begin{equation}
    \begin{aligned}
        &m^{1-loop}_8[12345678|12846573]=\int\frac{d^Dl}{(2\pi)^D}\frac{1}{l^2}\lim_{forward}\int d\Omega\\
        \{&[1-\frac{z_{+1}z_{-2}}{z_{-1}z_{+2}}-\frac{z_{18}z_{+-}}{z_{1-}z_{+8}}\times\frac{z_{18}z_{-2}}{z_{12}z_{-8}}-\frac{z_{-+}z_{23}}{z_{-3}z_{2+}}\times\frac{z_{+1}z_{23}}{z_{+3}z_{21}}]PT(1+-2345678)PT(1+-2846573)\\
        +&[1-\frac{z_{+2}z_{-3}}{z_{+3}z_{-2}}-\frac{z_{21}z_{+-}}{z_{2-}z_{+1}}\times\frac{z_{21}z_{-3}}{z_{23}z_{-1}}]PT(12+-345678)PT(12+-846573)\\
        +&[1-\frac{z_{+8}z_{-1}}{z_{+1}z_{-8}}-\frac{z_{-+}z_{12}}{z_{-2}z_{1+}}\times\frac{z_{+8}z_{12}}{z_{+2}z_{18}}]PT(12345678+-)PT(12846573+-)\\
        +&[1-\frac{z_{+2}z_{-3}}{z_{+3}z_{-2}}]PT(12+-345678)PT(128+-46573)\\
        +&[1-\frac{z_{+3}z_{-4}}{z_{+4}z_{-3}}]PT(123+-45678)PT(12+-846573)\\
        +&[1-\frac{z_{+3}z_{-4}}{z_{+4}z_{-3}}-\frac{z_{-+}z_{45}}{z_{-5}z_{4+}}\times\frac{z_{+3}z_{45}}{z_{+5}z_{43}}]PT(123+-45678)PT(128+-46573)\\
        +&[1-\frac{z_{+7}z_{-8}}{z_{+8}z_{-7}}-\frac{z_{76}z_{+-}}{z_{7-}z_{+6}}\times\frac{z_{76}z_{-8}}{z_{78}z_{-6}}]PT(1234567+-8)PT(1284657+-3)\\
        +&[1-\frac{z_{+7}z_{-8}}{z_{+8}z_{-7}}]PT(1234567+-8)PT(12846573+-)\\
        +&[1-\frac{z_{+8}z_{-1}}{z_{+1}z_{-8}}]PT(12345678+-)PT(1284657+-3)\\
        +&[1-\frac{z_{+4}z_{-5}}{z_{+5}z_{-4}}-\frac{z_{43}z_{+-}}{z_{4-}z_{+3}}\times\frac{z_{43}z_{-5}}{z_{45}z_{-3}}]PT(1234+-5678)PT(1284+-6573)\\
        +&[1-\frac{z_{+6}z_{-7}}{z_{+7}z_{-6}}-\frac{z_{-+}z_{78}}{z_{-8}z_{7+}}\times\frac{z_{+6}z_{78}}{z_{+8}z_{76}}]PT(123456+-78)PT(128465+-73)\}.
    \end{aligned}
    \label{eq:reslut of 12846573}
\end{equation}
Compared it with construction \eqref{eq:HY loop to tree} with $64$ terms, the difference it
exactly those massless bubbles.

\subsubsection{The second example}

The second example is  $m^{1-loop}_9[123456789|127893654]$ with its   effective tree Feynman diagram given by
\begin{equation}
    \includegraphics[scale=0.3]{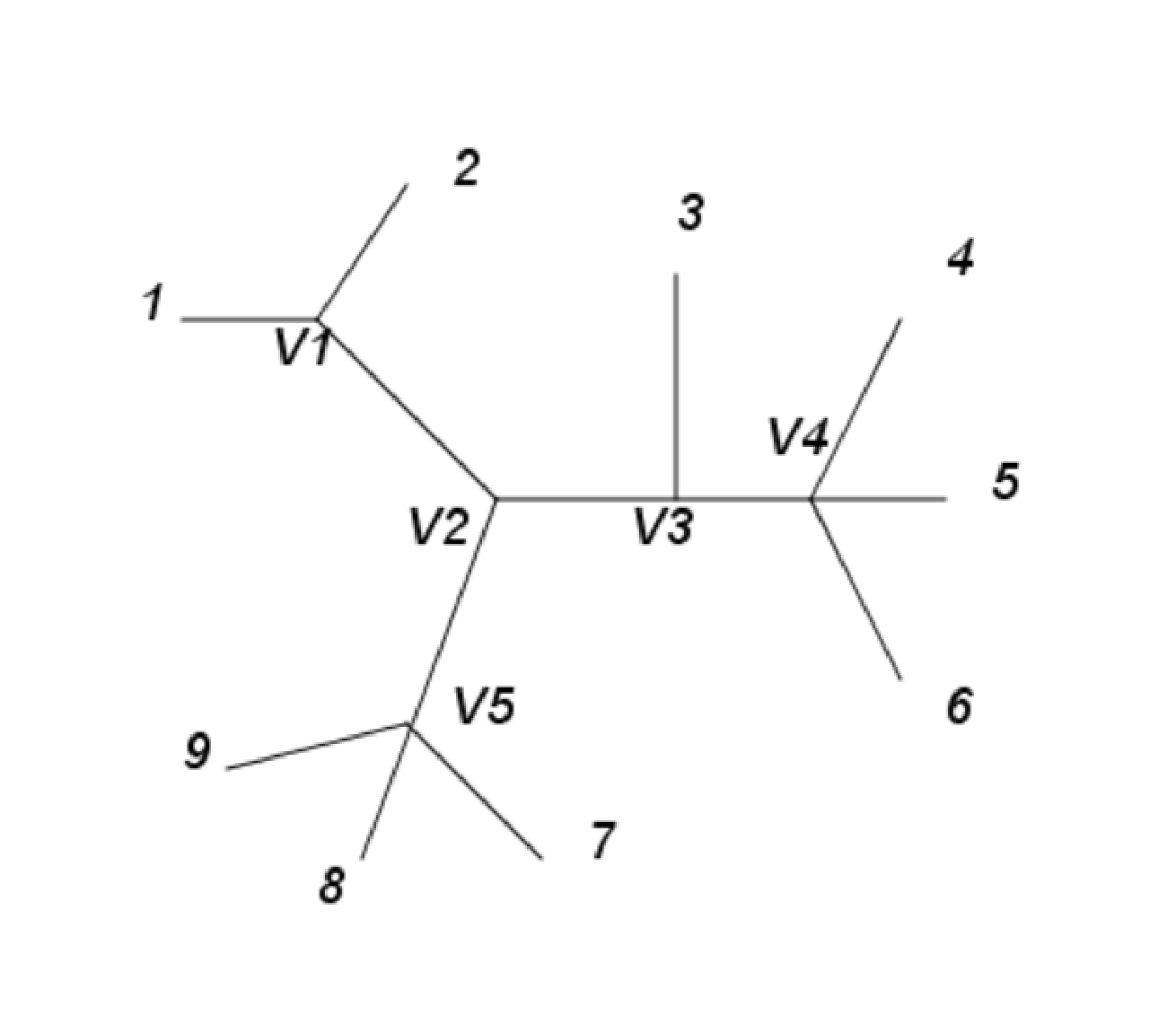}
\end{equation}
where $V_1, V_3,V_5$ are $\cal{SO}$-type and $V_2, V_4$, $\cal{RO}$-type.
For $V_1$, we have
\begin{equation}
    \begin{aligned}
        &(1-{\cal{P}}^{+,1}_{-,2})PT[(1)+-(2)(3456789)]PT[(1)+-(2)(7893654)]\\
        &(1-{\cal{P}}^{+,2}_{-,3})PT[(1)(2)+-(3456789)]PT[(1)(2)+-(7893654)]\\
        &(1-{\cal{P}}^{+,9}_{-,1})PT[(1)(2)(3456789)+-]PT[(1)(2)(7893654)+-].
    \end{aligned}
    \label{eq:V_1}
\end{equation}
For $V_2$, we have no contribution.
For $V_3$ we have:
\begin{equation}
    \begin{aligned}
        &(1-{\cal{P}}^{+,2}_{-,3})PT[(78912)+-(3)(456)]PT[(12789)+-(3)(654)]\\
        &(1-{\cal{P}}^{+,3}_{-,4})PT[(78912)(3)+-(456)]PT[(12789)(3)+-(654)]\\
        &(1-{\cal{P}}^{+,6}_{-,7})PT[(78912)(3)(456)+-]PT[(12789)(3)(654)+-].
    \end{aligned}
    \label{eq:V_3}
\end{equation}
For $V_4$, we have:
\begin{equation}
    \begin{aligned}
       &(1-{\cal{P}}^{+,3}_{-,4})PT[(789123)+-(4)(5)(6)]PT[(127893)(6)+-(5)(4)]\\
       &(1-{\cal{P}}^{+,4}_{-,5})PT[(789123)(4)+-(5)(6)]PT[(127893)+-(6)(5)(4)]\\
       &(1-{\cal{P}}^{+,5}_{-,6})PT[(789123)(4)(5)+-(6)]PT[(127893)(6)(5)(4)+-]\\
       &(1-{\cal{P}}^{+,6}_{-,7})PT[(789123)(4)(5)(6)+-]PT[(127893)(6)(5)+-(4)].
    \end{aligned}
    \label{eq:V_4}
\end{equation}
For $V_5$, we have:
\begin{equation}
    \begin{aligned}
       &(1-{\cal{P}}^{+,6}_{-,7})PT[(123456)+-(7)(8)(9)]PT[(365412)+-(7)(8)(9)]\\
       &(1-{\cal{P}}^{+,7}_{-,8})PT[(123456)(7)+-(8)(9)]PT[(365412)(7)+-(8)(9)]\\
       &(1-{\cal{P}}^{+,8}_{-,9})PT[(123456)(7)(8)+-(9)]PT[(365412)(7)(8)+-(9)]\\
       &(1-{\cal{P}}^{+,9}_{-,1})PT[(123456)(7)(8)(9)+-]PT[(365412)(7)(8)(9)+-].
    \end{aligned}
    \label{eq:V_5}
\end{equation}
Assembling above, we have 14 terms in total. By applying picking operator $\cal P$ to massless bubbles, we obtain:
\bea
 &  &m^{1-loop}_9[123456789|127893654]=\int\frac{d^Dl}{(2\pi)^D}\frac{1}{l^2}\lim_{forward}\int d\Omega \nn
& &        \left\{[1-\frac{z_{+1}z_{-2}}{z_{-1}z_{+2}}-\frac{z_{19}z_{+-}}{z_{1-}z_{+9}}\times\frac{z_{19}z_{-2}}
{z_{12}z_{-9}}-\frac{z_{-+}z_{23}}{z_{-3}z_{2+}}\times\frac{z_{+1}z_{23}}{z_{+3}z_{21}}]PT(1+-23456789)
PT(1+-27893654)\right.\nn
&+&[1-\frac{z_{+2}z_{-3}}{z_{+3}z_{-2}}-\frac{z_{21}z_{+-}}{z_{2-}z_{+1}}\times\frac{z_{21}z_{-3}}
{z_{23}z_{-1}}]PT(12+-3456789)PT(12+-7893654)\nn
&+&[1-\frac{z_{+9}z_{-1}}{z_{+1}z_{-9}}-\frac{z_{-+}z_{12}}{z_{-2}z_{1+}}\times\frac{z_{+9}z_{12}}{z_{+2}
z_{19}}]PT(123456789+-)PT(127893654+-)\nn
&+&[1-\frac{z_{+2}z_{-3}}{z_{+3}z_{-2}}-\frac{z_{-+}z_{34}}{z_{-4}z_{3+}}\times\frac{z_{+9}z_{34}}{z_{+4}
z_{39}}]PT(12+-3456789)PT(12789+-3654)\nn
&+&[1-\frac{z_{+3}z_{-4}}{z_{+4}z_{-3}}-\frac{z_{32}z_{+-}}{z_{3-}z_{+2}}\times\frac{z_{32}z_{-4}}
{z_{34}z_{-2}}]PT(123+-456789)PT(127893+-654)\nn
&+&[1-\frac{z_{+6}z_{-7}}{z_{+7}z_{-6}}]PT(123456+-789)PT(127893654+-)\nn
&+&[1-\frac{z_{+3}z_{-4}}{z_{+4}z_{-3}}]PT(123+-456789)PT(1278936+-54)\nn
&+&[1-\frac{z_{+4}z_{-5}}{z_{+5}z_{-4}}]PT(1234+-56789)PT(127893+-654)\nn
& +&[1-\frac{z_{+5}z_{-6}}{z_{+6}z_{-5}}]PT(12345+-6789)PT(127893654+-)\nn
&+&[1-\frac{z_{+6}z_{-7}}{z_{+7}z_{-6}}]PT(123456+-789)PT(12789365+-4)\nn
&+&[1-\frac{z_{+6}z_{-7}}{z_{+7}z_{-6}}-\frac{z_{-+}z_{78}}{z_{-8}z_{7+}}
\times\frac{z_{+6}z_{78}}{z_{+8}z_{76}}]PT(123456+-789)PT(12+-7893654)\nn
&+&[1-\frac{z_{+7}z_{-8}}{z_{-8}z_{+7}}-\frac{z_{76}z_{+-}}{z_{7-}z_{+6}}\times\frac{z_{76}z_{-8}}{z_{78}z_{-6}}
-\frac{z_{-+}z_{89}}{z_{-9}z_{8+}}\times\frac{z_{+7}z_{89}}{z_{+9}z_{87}}]PT(1234567+-89)PT(127+-893654)\nn
&+&[1-\frac{z_{+8}z_{-9}}{z_{-9}z_{+8}}-\frac{z_{87}z_{+-}}{z_{8-}z_{+7}}\times
\frac{z_{87}z_{-9}}{z_{89}z_{-7}}-\frac{z_{-+}z_{91}}{z_{-1}z_{9+}}\times
\frac{z_{+8}z_{91}}{z_{+1}z_{98}}]PT(12345678+-9)PT(1278+-93654)\nn
&+&\left.[1-\frac{z_{+9}z_{-1}}{z_{-9}z_{+1}}-\frac{z_{98}z_{+-}}{z_{9-}z_{+8}}\times\frac{z_{98}z_{-1}}
{z_{91}z_{-8}}]
PT(123456789+-)PT(12789+-3654)\right\}~~~.
   \label{eq:reslut of 127893654}
\eea
Compared it with construction \eqref{eq:HY loop to tree} with $81$ terms, the difference is exactly those external bubbles.

\section{Conclusion}

In this paper, we have focused on the construction of one-loop CHY-integrands for
 bi-adjoint scalar theory. Our
construction is different from the one given in
\cite{Baadsgaard:2015hia,He:2015yua} by the treatment of
singularities related to  the forward limit. Now it is well known
that one can construct one-loop amplitudes from tree's by taking the
forward limits. However, in the forward limit, some singularities as
well as extra terms will appear, such as the tadpoles and massless
bubbles, which are not considered for the standard amputated
one-loop Feynman diagrams. Different treatments of singularities and
extra terms as well as the singular solutions from scattering
equations will give different construction of one-loop
CHY-integrands, although all of them are in the same equivalent
class.   Our new strategy in this paper is to use the picking pole
operators to remove the singularities and extra terms explicitly.
More explicitly, by exploiting the concept of effective Feynman
diagrams, we show how to find corresponding tree-level diagrams and
how to remove singular tadpoles and massless bubbles by using the
cross ratio factor defined in \cite{Feng:2016nrf}.

The reason we investigate the new strategy is to hope that such a
method can be generalized to more general theories appearing in the
CHY formalism, such as Yang-Mills theories, NLSM etc. It has been
established that all theories can be expanded to bi-adjoint scaler
theory \cite{Bjerrum-Bohr:2016axv}. For example,  in
\cite{Fu:2017uzt} the expansion of Pfaffian to the combination of PT
factors has been presented. Thus Yang-Mills CHY-integrand can be
written as the sum of bi-adjoint scalar theories. Using results in
this paper, we  remove all singularities and extra terms for each
bi-adjoint CHY-integrand in the expansion. Now we end up a sum
without any singularities by the brute force way, but is it the
right result? Checking this idea is the future project we are
working on. Furthermore, even above thought works, expansion to PT
factors is very complicated, thus looking for a better way to remove
singularities and extra terms along the line of this paper is also
intriguing questions to ask.

\section*{Acknowledgments}

We would like to thank E.~Y.~Yuan for discussion and reading of
draft. This work is supported by Qiu-Shi Funding and Chinese NSF
funding under Grant No.11575156, No.11935013.


\begin{thebibliography}{99}

\bibitem{Cachazo:2013gna}
  F.~Cachazo, S.~He and E.~Y.~Yuan,
  ``Scattering equations and Kawai-Lewellen-Tye orthogonality,''
  Phys.\ Rev.\ D {\bf 90}, no. 6, 065001 (2014)
  doi:10.1103/PhysRevD.90.065001
  [arXiv:1306.6575 [hep-th]].

\bibitem{Cachazo:2013hca}
  F.~Cachazo, S.~He and E.~Y.~Yuan,
  ``Scattering of Massless Particles in Arbitrary Dimensions,''
  Phys.\ Rev.\ Lett.\  {\bf 113}, no. 17, 171601 (2014)
  doi:10.1103/PhysRevLett.113.171601
  [arXiv:1307.2199 [hep-th]].

\bibitem{Cachazo:2013iea}
  F.~Cachazo, S.~He and E.~Y.~Yuan,
  ``Scattering of Massless Particles: Scalars, Gluons and Gravitons,''
  JHEP {\bf 1407}, 033 (2014)
  doi:10.1007/JHEP07(2014)033
  [arXiv:1309.0885 [hep-th]].

\bibitem{Cachazo:2014nsa}
  F.~Cachazo, S.~He and E.~Y.~Yuan,
  ``Einstein-Yang-Mills Scattering Amplitudes From Scattering Equations,''
  JHEP {\bf 1501}, 121 (2015)
  doi:10.1007/JHEP01(2015)121
  [arXiv:1409.8256 [hep-th]].

\bibitem{Cachazo:2014xea}
  F.~Cachazo, S.~He and E.~Y.~Yuan,
  ``Scattering Equations and Matrices: From Einstein To Yang-Mills, DBI and NLSM,''
  JHEP {\bf 1507}, 149 (2015)
  doi:10.1007/JHEP07(2015)149
  [arXiv:1412.3479 [hep-th]].

\bibitem{Dolan:2013isa}
  L.~Dolan and P.~Goddard,
  ``Proof of the Formula of Cachazo, He and Yuan for Yang-Mills Tree Amplitudes in Arbitrary Dimension,''
  JHEP {\bf 1405}, 010 (2014)
  doi:10.1007/JHEP05(2014)010
  [arXiv:1311.5200 [hep-th]].



\bibitem{Dolan:2014ega}
  L.~Dolan and P.~Goddard,
  ``The Polynomial Form of the Scattering Equations,''
  JHEP {\bf 1407}, 029 (2014)
  doi:10.1007/JHEP07(2014)029
  [arXiv:1402.7374 [hep-th]].

\bibitem{Kalousios:2015fya}
  C.~Kalousios,
  ``Scattering equations, generating functions and all massless five point tree amplitudes,''
  JHEP {\bf 1505}, 054 (2015)
  doi:10.1007/JHEP05(2015)054
  [arXiv:1502.07711 [hep-th]].



\bibitem{Huang:2015yka}
  R.~Huang, J.~Rao, B.~Feng and Y.~H.~He,
  ``An Algebraic Approach to the Scattering Equations,''
  JHEP {\bf 1512}, 056 (2015)
  doi:10.1007/JHEP12(2015)056
  [arXiv:1509.04483 [hep-th]].

\bibitem{Sogaard:2015dba}
  M.~Sgaard and Y.~Zhang,
  ``Scattering Equations and Global Duality of Residues,''
  Phys.\ Rev.\ D {\bf 93}, no. 10, 105009 (2016)
  doi:10.1103/PhysRevD.93.105009
  [arXiv:1509.08897 [hep-th]].

\bibitem{Dolan:2015iln}
  L.~Dolan and P.~Goddard,
  ``General Solution of the Scattering Equations,''
  JHEP {\bf 1610}, 149 (2016)
  doi:10.1007/JHEP10(2016)149
  [arXiv:1511.09441 [hep-th]].

\bibitem{Cardona:2015eba}
  C.~Cardona and C.~Kalousios,
  ``Comments on the evaluation of massless scattering,''
  JHEP {\bf 1601}, 178 (2016)
  doi:10.1007/JHEP01(2016)178
  [arXiv:1509.08908 [hep-th]].

\bibitem{Cardona:2015ouc}
  C.~Cardona and C.~Kalousios,
  ``Elimination and recursions in the scattering equations,''
  Phys.\ Lett.\ B {\bf 756}, 180 (2016)
  doi:10.1016/j.physletb.2016.03.003
  [arXiv:1511.05915 [hep-th]].


\bibitem{Cachazo:2015nwa}
  F.~Cachazo and H.~Gomez,
  ``Computation of Contour Integrals on ${\cal M}_{0,n}$,''
  JHEP {\bf 1604}, 108 (2016)
  doi:10.1007/JHEP04(2016)108
  [arXiv:1505.03571 [hep-th]].

\bibitem{Cachazo:2019aby}
  F.~Cachazo, K.~Yeats and S.~Yusim,
  ``Compatible Cycles and CHY Integrals,''
  arXiv:1907.12661 [math-ph].

\bibitem{Baadsgaard:2015voa}
  C.~Baadsgaard, N.~E.~J.~Bjerrum-Bohr, J.~L.~Bourjaily and P.~H.~Damgaard,
  ``Integration Rules for Scattering Equations,''
  JHEP {\bf 1509}, 129 (2015)
  doi:10.1007/JHEP09(2015)129
  [arXiv:1506.06137 [hep-th]].


\bibitem{Baadsgaard:2015ifa}
  C.~Baadsgaard, N.~E.~J.~Bjerrum-Bohr, J.~L.~Bourjaily and P.~H.~Damgaard,
  ``Scattering Equations and Feynman Diagrams,''
  JHEP {\bf 1509}, 136 (2015)
  doi:10.1007/JHEP09(2015)136
  [arXiv:1507.00997 [hep-th]].

\bibitem{Baadsgaard:2015hia}
  C.~Baadsgaard, N.~E.~J.~Bjerrum-Bohr, J.~L.~Bourjaily, P.~H.~Damgaard and B.~Feng,
  ``Integration Rules for Loop Scattering Equations,''
  JHEP {\bf 1511}, 080 (2015)
  doi:10.1007/JHEP11(2015)080
  [arXiv:1508.03627 [hep-th]].

\bibitem{Cardona:2016gon}
  C.~Cardona, B.~Feng, H.~Gomez and R.~Huang,
  ``Cross-ratio Identities and Higher-order Poles of CHY-integrand,''
  JHEP {\bf 1609}, 133 (2016)
  doi:10.1007/JHEP09(2016)133
  [arXiv:1606.00670 [hep-th]].

\bibitem{Lam:2016tlk}
  C.~S.~Lam and Y.~P.~Yao,
  ``Evaluation of the Cachazo-He-Yuan gauge amplitude,''
  Phys.\ Rev.\ D {\bf 93}, no. 10, 105008 (2016)
  doi:10.1103/PhysRevD.93.105008
  [arXiv:1602.06419 [hep-th]].

\bibitem{Adamo:2013tsa}
  T.~Adamo, E.~Casali and D.~Skinner,
  ``Ambitwistor strings and the scattering equations at one loop,''
  JHEP {\bf 1404}, 104 (2014)
  doi:10.1007/JHEP04(2014)104
  [arXiv:1312.3828 [hep-th]].

\bibitem{Casali:2014hfa}
  E.~Casali and P.~Tourkine,
  ``Infrared behaviour of the one-loop scattering equations and supergravity integrands,''
  JHEP {\bf 1504}, 013 (2015)
  doi:10.1007/JHEP04(2015)013
  [arXiv:1412.3787 [hep-th]].

\bibitem{Adamo:2015hoa}
  T.~Adamo and E.~Casali,
  ``Scattering equations, supergravity integrands, and pure spinors,''
  JHEP {\bf 1505}, 120 (2015)
  doi:10.1007/JHEP05(2015)120
  [arXiv:1502.06826 [hep-th]].

\bibitem{Geyer:2015bja}
  Y.~Geyer, L.~Mason, R.~Monteiro and P.~Tourkine,
  ``Loop Integrands for Scattering Amplitudes from the Riemann Sphere,''
  Phys.\ Rev.\ Lett.\  {\bf 115}, no. 12, 121603 (2015)
  doi:10.1103/PhysRevLett.115.121603
  [arXiv:1507.00321 [hep-th]].



\bibitem{Geyer:2015jch}
  Y.~Geyer, L.~Mason, R.~Monteiro and P.~Tourkine,
  ``One-loop amplitudes on the Riemann sphere,''
  JHEP {\bf 1603}, 114 (2016)
  doi:10.1007/JHEP03(2016)114
  [arXiv:1511.06315 [hep-th]].


\bibitem{He:2015yua}
  S.~He and E.~Y.~Yuan,
  ``One-loop Scattering Equations and Amplitudes from Forward Limit,''
  Phys.\ Rev.\ D {\bf 92}, no. 10, 105004 (2015)
  doi:10.1103/PhysRevD.92.105004
  [arXiv:1508.06027 [hep-th]].

\bibitem{Cachazo:2015aol}
  F.~Cachazo, S.~He and E.~Y.~Yuan,
  ``One-Loop Corrections from Higher Dimensional Tree Amplitudes,''
  JHEP {\bf 1608}, 008 (2016)
  doi:10.1007/JHEP08(2016)008
  [arXiv:1512.05001 [hep-th]].

\bibitem{Zlotnikov:2016wtk}
  M.~Zlotnikov,
  ``Polynomial reduction and evaluation of tree- and loop-level CHY amplitudes,''
  JHEP {\bf 1608}, 143 (2016)
  doi:10.1007/JHEP08(2016)143
  [arXiv:1605.08758 [hep-th]].

\bibitem{Cardona:2016wcr}
  C.~Cardona and H.~Gomez,
  ``CHY-Graphs on a Torus,''
  JHEP {\bf 1610}, 116 (2016)
  doi:10.1007/JHEP10(2016)116
  [arXiv:1607.01871 [hep-th]].

\bibitem{He:2016mzd}
  S.~He and O.~Schlotterer,
  ``New Relations for Gauge-Theory and Gravity Amplitudes at Loop Level,''
  Phys.\ Rev.\ Lett.\  {\bf 118}, no. 16, 161601 (2017)
  doi:10.1103/PhysRevLett.118.161601
  [arXiv:1612.00417 [hep-th]].

\bibitem{Gomez:2016cqb}
  H.~Gomez, S.~Mizera and G.~Zhang,
  ``CHY Loop Integrands from Holomorphic Forms,''
  JHEP {\bf 1703}, 092 (2017)
  doi:10.1007/JHEP03(2017)092
  [arXiv:1612.06854 [hep-th]].

\bibitem{Gomez:2017lhy}
  H.~Gomez,
  ``Quadratic Feynman Loop Integrands From Massless Scattering Equations,''
  Phys.\ Rev.\ D {\bf 95}, no. 10, 106006 (2017)
  doi:10.1103/PhysRevD.95.106006
  [arXiv:1703.04714 [hep-th]].



\bibitem{Gomez:2017cpe}
  H.~Gomez, C.~Lopez-Arcos and P.~Talavera,
  ``One-loop Parke-Taylor factors for quadratic propagators from massless scattering equations,''
  JHEP {\bf 1710}, 175 (2017)
  doi:10.1007/JHEP10(2017)175
  [arXiv:1707.08584 [hep-th]].

\bibitem{Geyer:2017ela}
  Y.~Geyer and R.~Monteiro,
  ``Gluons and gravitons at one loop from ambitwistor strings,''
  JHEP {\bf 1803}, 068 (2018)
  doi:10.1007/JHEP03(2018)068
  [arXiv:1711.09923 [hep-th]].

\bibitem{Ahmadiniaz:2018nvr}
  N.~Ahmadiniaz, H.~Gomez and C.~Lopez-Arcos,
  JHEP {\bf 1805}, 055 (2018)
  doi:10.1007/JHEP05(2018)055
  [arXiv:1802.00015 [hep-th]].

\bibitem{Agerskov:2019ryp}
  J.~Agerskov, N.~E.~J.~Bjerrum-Bohr, H.~Gomez and C.~Lopez-Arcos,
  ``Yang-Mills Loop Amplitudes from Scattering Equations,''
  arXiv:1910.03602 [hep-th].


\bibitem{Feng:2016nrf}
  B.~Feng,
  ``CHY-construction of Planar Loop Integrands of Cubic Scalar Theory,''
  JHEP {\bf 1605}, 061 (2016)
  doi:10.1007/JHEP05(2016)061
  [arXiv:1601.05864 [hep-th]].

\bibitem{Geyer:2016wjx}
  Y.~Geyer, L.~Mason, R.~Monteiro and P.~Tourkine,
  ``Two-Loop Scattering Amplitudes from the Riemann Sphere,''
  Phys.\ Rev.\ D {\bf 94}, no. 12, 125029 (2016)
  doi:10.1103/PhysRevD.94.125029
  [arXiv:1607.08887 [hep-th]].


\bibitem{Geyer:2018xwu}
  Y.~Geyer and R.~Monteiro,
  ``Two-Loop Scattering Amplitudes from Ambitwistor Strings: from Genus Two to the Nodal Riemann Sphere,''
  JHEP {\bf 1811}, 008 (2018)
  doi:10.1007/JHEP11(2018)008
  [arXiv:1805.05344 [hep-th]].

\bibitem{Geyer:2019hnn}
  Y.~Geyer, R.~Monteiro and R.~Stark-Muchao,
  ``Two-Loop Scattering Amplitudes: Double-Forward Limit and Colour-Kinematics Duality,''
  arXiv:1908.05221 [hep-th].


\bibitem{Huang:2016zzb}
  R.~Huang, B.~Feng, M.~x.~Luo and C.~J.~Zhu,
  ``Feynman Rules of Higher-order Poles in CHY Construction,''
  JHEP {\bf 1606}, 013 (2016)
  doi:10.1007/JHEP06(2016)013
  [arXiv:1604.07314 [hep-th]].

\bibitem{Huang:2017ydz}
  R.~Huang, Y.~J.~Du and B.~Feng,
  ``Understanding the Cancelation of Double Poles in the Pfaffian of CHY-formulism,''
  JHEP {\bf 1706}, 133 (2017)
  doi:10.1007/JHEP06(2017)133
  [arXiv:1702.05840 [hep-th]].


\bibitem{Huang:2018zfi}
  R.~Huang, F.~Teng and B.~Feng,
  ``Permutation in the CHY-Formulation,''
  Nucl.\ Phys.\ B {\bf 932}, 323 (2018)
  doi:10.1016/j.nuclphysb.2018.05.014
  [arXiv:1801.08965 [hep-th]].

\bibitem{Baadsgaard:2015abc}
C.~Baadsgaard, Amplitudes from sting theory and CHY formalism,
  Master's thesis, Copenhagen University, 2015.
Available at
  http://discoverycenter.nbi.ku.dk/teaching/thesis\_page/.













\bibitem{Bjerrum-Bohr:2016juj}
N.~E.~J.~Bjerrum-Bohr, J.~L.~Bourjaily, P.~H.~Damgaard and B.~Feng,
'Analytic representations of Yang-Mills amplitudes,'
Nucl.\ Phys.\ B {\bf 913}, 964 (2016)
doi:10.1016/j.nuclphysb.2016.10.012
[arXiv:1605.06501 [hep-th]].



\bibitem{Bjerrum-Bohr:2016axv}
  N.~E.~J.~Bjerrum-Bohr, J.~L.~Bourjaily, P.~H.~Damgaard and B.~Feng,
  ``Manifesting Color-Kinematics Duality in the Scattering Equation Formalism,''
  JHEP {\bf 1609}, 094 (2016)
  doi:10.1007/JHEP09(2016)094
  [arXiv:1608.00006 [hep-th]].



\bibitem{Zhou:2017mfj}
  K.~Zhou, J.~Rao and B.~Feng,
  ``Derivation of Feynman Rules for Higher Order Poles Using Cross-ratio Identities in CHY Construction,''
  JHEP {\bf 1706}, 091 (2017)
  doi:10.1007/JHEP06(2017)091
  [arXiv:1705.04783 [hep-th]].


\bibitem{Fu:2017uzt}
  C.~H.~Fu, Y.~J.~Du, R.~Huang and B.~Feng,
  ``Expansion of Einstein-Yang-Mills Amplitude,''
  JHEP {\bf 1709}, 021 (2017)
  doi:10.1007/JHEP09(2017)021
  [arXiv:1702.08158 [hep-th]].

\end{thebibliography}
\end{document}